%% file: hh202.tex
\documentclass[useAMS,usenatbib]{mn2e}

\usepackage[dvips]{graphicx}
\usepackage{amssymb}
\usepackage{amsmath}
\usepackage{amstext}

\newcommand{\adfo}{${\rm ADF(O^{2+})}$}

\newcommand{\te}{$T_{e}$}
\newcommand{\nel}{$n_{e}$}
\newcommand{\vhel}{$V_{hel}$}
\newcommand{\tf}{$t^{2}$}
\newcommand\ion[2]{#1~{\sc {#2}}\relax}        
\newcommand\ioni[2]{${\rm #1^{#2}}$}           

\newcommand{\cmc}{{\rm cm$^{-3}$}}
\newcommand{\kms}{{\rm km~s$^{-1}$}}

\newcommand{\neb}{$_{\rm neb}$}
\newcommand{\sho}{$_{\rm sh}$}
\newcommand{\hh}{HH~202}
\newcommand{\hii}{H~{\sc ii}}
\newcommand{\hi}{H~{\sc i}}
\newcommand{\ld}{$\lambda$}

\title[Echelle spectrophotometry of \hh]{Properties of the ionized gas in \hh. II: Results 
                                         from echelle spectrophotometry with UVES%
			       \thanks{Based on observations collected at the European Southern 
			       Observatory, Chile, proposal number ESO 70.C-0008(A).}}
\author[A. Mesa-Delgado et al.]
       {A. Mesa-Delgado$^1$\thanks{E-mail: amd@iac.es}, 
        C. Esteban$^1$, J. Garc{\'{\i}}a-Rojas$^2$, V. Luridiana$^3$, M. Bautista$^4$,   
	\newauthor 
	M. Rodr{\'{\i}}guez$^5$, L. L\'opez-Mart{\'{\i}}n$^1$ and M. Peimbert$^2$\\
	$^1$Instituto de Astrof\'\i sica de Canarias, E-38200 La Laguna, Tenerife, Spain \\
        $^2$Instituto de Astronom\'\i a, UNAM, Apdo. Postal 70-264, 04510 M\'exico D.F., Mexico\\
        $^3$Instituto de Astrof\'\i sica de Andaluc\'\i a (CSIC), Apdo. Correos 3004, E-18080 
	    Granada, Spain\\
        $^4$Deparment of Physics, Virginia Polytechnic and State University, Blacksburg, VA 
	    24061, USA\\
        $^5$Instituto Nacional de Astrof\'\i sica, \'Optica y Electr\'onica INAOE, Apdo. Postal 51 
	    y 216, 7200 Puebla, Pue., Mexico\\}

\begin{document}

\date{Accepted 2009 January 23.  Received 2008 December 23; in original form 2008 October 24}
\pagerange{\pageref{firstpage}--\pageref{lastpage}} \pubyear{2008}

\maketitle
\label{firstpage}

\begin{abstract}
 We present results of deep echelle spectrophotometry of the brightest knot of the Herbig-Haro 
 object \hh\ in the Orion Nebula --\hh-S-- using the Ultraviolet Visual Echelle Spectrograph (UVES) 
 in the spectral range from 3100 to 10400 \AA. The high spectral resolution of the observations has 
 permitted to separate the component associated with the ambient gas from that associated with the 
 gas flow. We derive electron densities and temperatures from different diagnostics for both 
 components, as well as the chemical abundances of several ions and elements from collisionally 
 excited lines, including the first determinations of \ioni{Ca}{+} and \ioni{Cr}{+} abundances in 
 the Orion Nebula. We also calculate the \ioni{He}{+}, \ioni{C}{2+}, \ioni{O}{+} and \ioni{O}{2+} 
 abundances from recombination lines. The difference between the \ioni{O}{2+} abundances determined 
 from collisionally excited and recombination lines --the so-called abundance discrepancy factor-- 
 is 0.35 dex and 0.11 dex for the shock and nebular components, respectively. Assuming that the 
 abundance discrepancy is produced by spatial variations in the electron temperature, we derive 
 values of the temperature fluctuation parameter, \tf, of 0.050 and 0.016, for the shock and 
 nebular components, respectively. Interestingly, we obtain almost coincident \tf\ values for both 
 components from the analysis of the intensity ratios of \ion{He}{i} lines. We find significant 
 departures from case B predictions in the Balmer and Paschen flux ratios of lines of high 
 principal quantum number $n$. We analize the ionization structure of \hh-S, finding enough 
 evidence to conclude that the flow of \hh-S has compressed the ambient gas inside the nebula 
 trapping the ionization front. We measure a strong increase of the total abundances of nickel and 
 iron in the shock component, the abundance pattern and the results of photoionization models for 
 both components are consistent with the partial destruction of dust after the passage of the 
 shock wave in \hh-S. 
\end{abstract}


\section{Introduction} \label{intro}
\begin{figure*}
   \centering
   \includegraphics[scale=0.7]{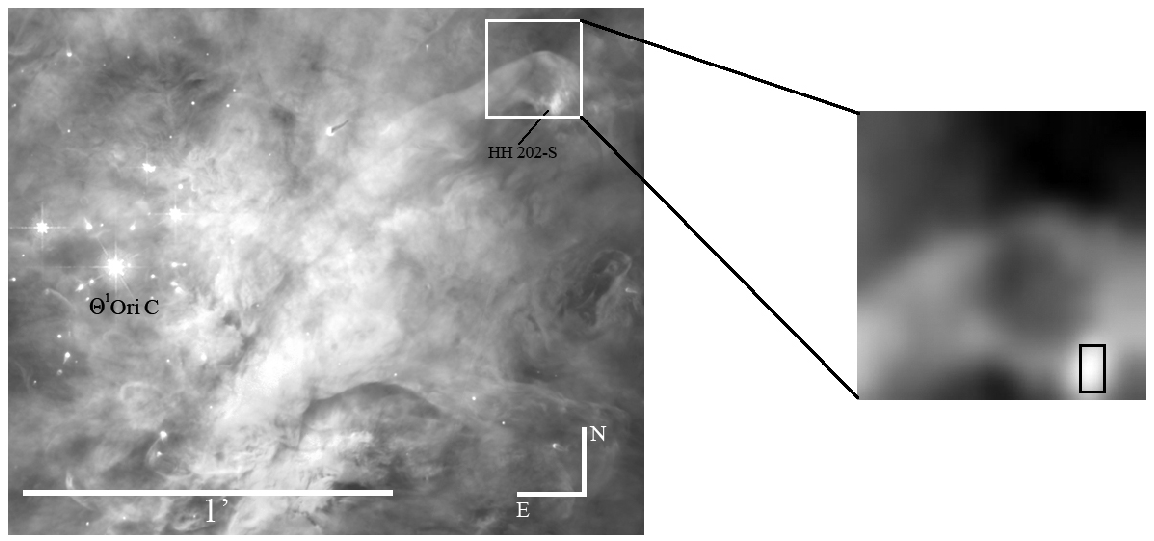} 
   \caption{$HST$ image of the central part of the Orion Nebula combined with WFPC2 images in 
            different filters \citep{odellwong96}. The white square corresponds to the field of 
	    view of the Potsdam Multi-Aperture Spectrograph (PMAS) and the separate close up image 
	    at the right shows the rebinned H$\alpha$ map presented in Paper I. Inside of this box, 
	    the black rectangle indicates the slit position and the area covered by the UVES 
	    spectrum analyzed in this paper (1\farcs5$\times$2\farcs5).}
   \label{f1}
  \end{figure*}

 \hh\ is one of the brightest and most conspicuous Herbig-Haro (HH) objects of the Orion Nebula. 
 It was discovered by \cite{cantoetal80}. The origin of this outflow is not clear, though the 
 radial velocity and proper motion studies suggest that this object forms a great complex together 
 with HH~203, 204, 269, 529, 528 and 625, with a common origin in one or more sources embedded 
 within the Orion Molecular Cloud 1 South (OMC 1S) \citep[see][]{rosadoetal02,odelldoi03,%
 odellhenney08}. Recently, \cite{henneyetal07} have summarized the main characteristics of these 
 outflows and an extensive study of their kinematics can be found in \cite{garciadiazetal08}. 
 \hh\ shows a wide parabolic form with several bright knots of which \hh-S is the brightest one 
 (see Figure~\ref{f1}).\\ 
 The kinematic properties of \hh-S have been studied by means of high-spectral resolution 
 spectroscopy by several authors. \cite{doietal04} have found a radial velocity of $-$39$\pm$2 
 \kms, in agreement with previous results by \cite{meaburn86} and \cite{odelletal91}. 
 \cite{odellhenney08} have determined a tangential velocity of 59$\pm$8 \kms, which is in agreement 
 with previous determinations by \cite{odelldoi03}. \cite{odellhenney08} have calculated a 
 spatial velocity of $89$ \kms\ and an angle of the velocity vector of 48$\rm ^o$ with respect to 
 the plane of the sky, similar to the values found by \cite{henneyetal07}. Imaging studies by 
 \cite{odelletal97} with the Hubble Space Telescope ($HST$) of the HH objects in the Orion Nebula 
 show an extended [\ion{O}{iii}] emission in \hh\ and strong [\ion{O}{iii}] emission in \hh-S. This 
 fact, together with the closeness of \hh\ to the main ionization source of the Orion Nebula, 
 $\theta^1$ Ori C, indicate that the excitation of the ionized gas is dominated by photoionization 
 in \hh-S, though the observed radial velocities imply that some shocked gas can be mixed in the 
 region \citep{cantoetal80}. Photoionization-dominated flows are a minority in the inventory 
 of HH objects, which are typically excited by shocks. This kind of HH objects is also known as 
 ``irradiated jets" \citep{reipurthetal98}, since they are excited by the UV radiation from nearby 
 massive stars. Irradiated jets have been found in the Orion Nebula 
 \citep[$e.g.$][]{ballyreipurth01,ballyetal06,odelletal97}, the Pelican Nebula 
 \citep{ballyreipurth03}, the Carina Nebula \citep{smithetal04}, NGC~1333 
 \citep{ballyetal06} and the Trifid Nebula \citep{cernicharoetal98,reipurthetal98}.\\
 \cite{mesadelgadoetal08a} have obtained the spatial distributions of the physical conditions and 
 the ionic abundances in the Orion Nebula using long-slit spectroscopy at spatial scales of 
 1\farcs2. The goal of that work was to study the possible correlations between the local 
 structures observed in the Orion Nebula --HH objects, proplyds, ionization fronts-- and the 
 abundance discrepancy (AD) that is found in \hii\ regions. The AD is a classical problem in the 
 study of ionized nebulae: the abundances of a given ion derived from recombination lines, RLs, are 
 often between 0.1 and 0.3 dex higher than those obtained from collisionally excited lines, CELs, 
 in \hii\ regions \cite[see][]{garciarojasesteban07,estebanetal04,tsamisetal03}. The difference 
 between those independent determinations of the abundance defines the abundance discrepancy 
 factor, ADF. The predictions of the temperature fluctuation paradigm proposed by \cite{peimbert67} 
 --and parametrized by the mean square of the spatial distribution of temperature, the \tf\ 
 parameter-- seem to account for the discrepancies observed in \hii\ regions \cite[see][]%
 {garciarojasesteban07}. A striking result found in the spatially-resolved study of 
 \cite{mesadelgadoetal08a} is that the ADF of \ioni{O}{2+}, \adfo, shows larger values at the 
 locations of HH objects as is the case of \hh. Using integral field spectroscopy with 
 intermediate-spectral resolution and a spatial resolution of 1$\arcsec\times$1$\arcsec$, 
 \cite{mesadelgadoetal08b} (hereinafter, Paper I) have mapped the emission line fluxes, the 
 physical properties and the \ioni{O}{2+} abundances derived from RLs and CELs of \hh. They have 
 found extended [\ion{O}{iii}] emission and higher values of the electron density and temperaure as 
 well as an enhanced \adfo\ in \hh-S, confirming the earlier results of 
 \cite{mesadelgadoetal08a}.\\
 HH~529 is another HH object that is photoionized by $\theta^1$ Ori C and shows similar 
 characteristics to those of \hh. \cite{blagraveetal06} have performed deep optical echelle 
 spectroscopy of that object with a 4m-class telescope and have detected and measured about 280 
 emission lines. Their high-spectral resolution spectroscopy allowed them to separate the kinematic 
 components associated with the ambient gas and with the flow. They have determined the physical 
 conditions and the ionic abundances of oxygen from CELs and RLs in both components. However, they 
 do not find high \adfo\ and \tf\ values in neither component. Another interesting result of 
 \cite{blagraveetal06} is that the ionization structure of HH~529 indicates that it is a 
 matter-bounded shock.\\
 Motivated by the results found by \cite{mesadelgadoetal08a}, inspired by the work of 
 \cite{blagraveetal06} and in order to complement the results presented in Paper I, we have 
 isolated the emission of the flow of \hh-S knot using high-spectral resolution spectroscopy, 
 presenting the first complete physical and chemical analysis of this knot.\\  
 In \S\ref{obsred} we describe the observations of \hh\ and the reduction procedure. In \S\ref{lir} 
 we describe the emission line measurements, identifications and the reddening correction, we also 
 compare our reddening determinations with those available in the literature. In \S\ref{resul} we 
 describe the determinations of the physical conditions, the chemical --ionic and total-- 
 abundances 
 and the ADF for \ioni{O}{+} and \ioni{O}{2+}. In \S\ref{discu} we discuss: a) some inconsistencies 
 found in the Balmer decrement of the lines of higher principal quantum number, b) the ionization 
 structure of \hh-S, c) the radial velocity pattern of the lines of each kinematic component, d) 
 the \tf\ parameter obtained from different methods and its possible relation with the ADF, and e) 
 the evidences of dust grain destruction in \hh-S. Finally, in \S\ref{conclu} we summarize our 
 main conclusions. 
  \begin{figure}
   \centering
   \includegraphics[scale=0.6]{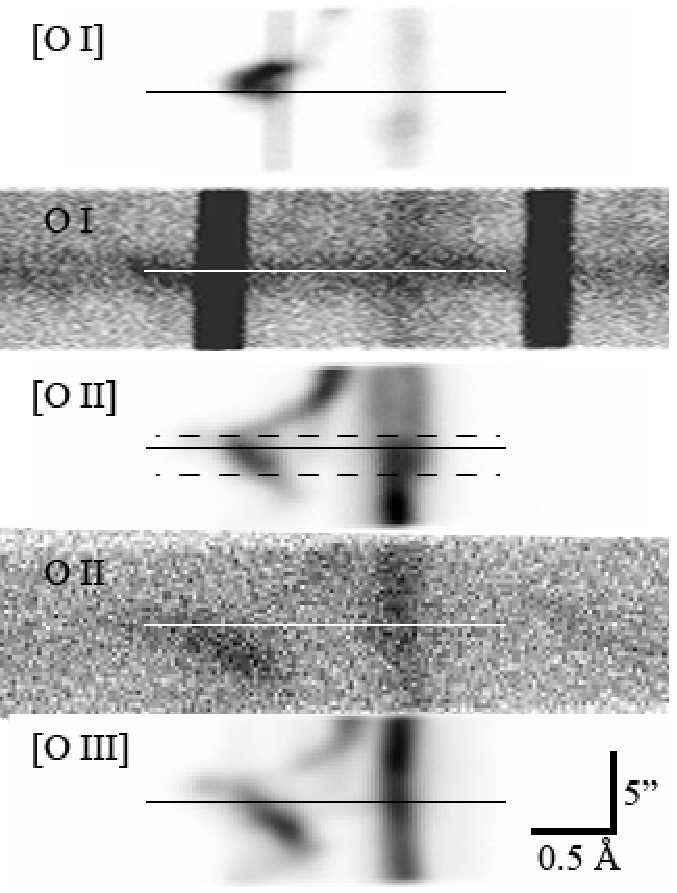} 
   \caption{Sections of the bidimensional UVES spectrum showing the spatio-kinematic profiles of 
            different oxygen lines: [\ion{O}{i}] 6300 \AA, \ion{O}{i} 7772 \AA, [\ion{O}{ii}] 3728 
	    \AA, \ion{O}{ii} 4649 \AA\ and [\ion{O}{iii}] 4959 \AA. Wavelength increases to the 
	    right and north points up. The solid straight lines in all the sections correspond to 
	    the slit center and the dashed lines --only represented in the case of the 
	    [\ion{O}{ii}] 3728 \AA\ profile-- correspond to the extracted area 2\farcs5 wide.}
   \label{f2}
  \end{figure}    	 
\section{Observations and Data Reduction} \label{obsred}
 \hh\ was observed on 2003 March 30 at Cerro Paranal Observatory (Chile), using the UT2 (Kueyen) 
 of the Very Large Telescope (VLT) with the Ultraviolet Visual Echelle Spectrograph 
 \citep[UVES,][]{dodoricoetal2000}. The standard settings of UVES were used covering the spectral 
 range from 3100 to 10400 \AA. Some narrow spectral ranges could not be observed. These are: 
 5783-5830 and 8540-8650 \AA, due to the physical separation between the CCDs of the detector 
 system of the red arm; and 10084-10088 and 10252-10259 \AA, because the last two orders of the 
 spectrum do not fit completely within the size of the CCD. Five individual exposures of 90 
 seconds --for the 3100-3900 and 4750-6800 \AA\ ranges-- and 270 seconds --for the 3800-5000 and 
 6700-10400 \AA\ ranges-- were added to obtain the final spectra. In addition, exposures of 5 and 
 10 seconds were taken to obtain good flux measurements --$i.e.$ non-saturated-- for the brightest 
 emission lines. The spectral resolution was $\lambda/\Delta\lambda\approx30000$. This high 
 spectral resolution enables us to separate two kinematic components: one corresponding to the 
 ambient gas --which we will call {\it nebular component} and whose emission mainly arises from 
 behind \hh\ and, therefore, could not entirely correspond to the pre-shock gas-- and another one 
 corresponding to the gas flow of the HH object, the post-shock gas, which we will call {\it shock 
 component}.\\ 
 The slit was oriented north-south and the atmospheric dispersion corrector (ADC) was used to keep 
 the same observed region within the slit regardless of the air mass value. The HH object was 
 observed between airmass values of 1.20 and 1.35. The average seeing during the observation was 
 0\farcs7. The slit width was set to 1\farcs5 as a compromise between the spectral resolution 
 needed and the desired signal-to-noise ratio of the spectra. The slit length was fixed to 
 10$\arcsec$. The one-dimensional spectra were extracted for an area of 1\farcs5$\times$2\farcs5. 
 This area covers the apex of \hh, the so-called knot \hh-S, as we can see in Figure~\ref{f1}. This 
 zone shows the maximum shift in velocity between the shock and nebular components (see 
 Figure~\ref{f2}) allowing us to appropriately separate and study the spectra of both kinematic 
 components.\\
 The spectra were reduced using the {\sc iraf}\footnote{{\sc iraf} is distributed by NOAO, which 
 is operated by AURA, under cooperative agreement with NSF.} echelle reduction package, following 
 the standard procedure of bias subtraction, aperture extraction, flatfielding, wavelength 
 calibration and flux calibration. The standard stars EG~247, C-32~9927 \citep{hamuyetal92,%
 hamuyetal94} and HD~49798 \citep{bohlinlindler92,turnsheketal90} were observed to perform the 
 flux calibration. The error of the absolute flux calibration was of the order of 3\%.
\section{Line measurements, identifications and reddening correction} \label{lir}
 Line fluxes were measured applying a double Gaussian profile fit procedure over the local 
 continuum. All these measurements were made with the {\sc splot} routine of {\sc iraf}.\\
 All line fluxes of a given spectrum have been normalized to a particular bright emission line 
 present in the common range of two consecutive spectra. For the bluest spectrum (3100-3900 \AA), 
 the reference line was H9 3835 \AA. For the range from 3800 to 5000 \AA, the reference line was 
 H$\beta$. In the case of the spectrum covering 4750-6800 \AA, the reference was [\ion{O}{iii}] 
 4959 \AA. Finally, for the reddest spectrum (6700-10400 \AA), the reference line was [\ion{S}{ii}] 
 6731 \AA. In order to produce a final homogeneous set of line flux ratios, all of them were 
 rescaled to the H$\beta$ flux. In the case of the bluest spectra the ratios were rescaled by the 
 H9/H$\beta$ ratio obtained from the 3800-5000 \AA\ range. The emission line ratios of the 
 4750-6800 \AA\ range were multipled by the [\ion{O}{iii}] 4959/H$\beta$ ratio measured in the 
 3800-5000 \AA\ range. In the case of the last spectral section, 6700-10400 \AA, the rescaling 
 factor was the [\ion{S}{ii}] 6731/H$\beta$ ratio obtained from the 4750-6800 \AA\ spectrum. All 
 rescaling factors were measured in the short exposure spectra in order to avoid the possible 
 saturation of the brightest emission lines. This process was done separately for both the nebular 
 and shock components.\\
 The spectral ranges present overlapping regions at the edges. The adopted flux of a line in the 
 overlapping region was obtained as the average of the values obtained in both spectra. A similar 
 procedure was considered in the case of lines present in two consecutive spectral orders of the 
 same spectral range. The average of both measurements was considered for the adopted value of 
 the line flux. In all cases, the differences in the line flux measured for the same line in 
 different orders and/or spectral ranges do not show systematic trends and are always within the 
 uncertainties.\\
 The identification and laboratory wavelengths of the lines were obtained following a previous work 
 on the Orion Nebula by \cite{estebanetal04}, the compilations by \cite{moore45} and The Atomic 
 Line List v2.04\footnote{Webpage at: http://www.pa.uky.edu/$\sim$peter/atomic/}. The 
  \begin{figure}
   \centering
   \includegraphics[scale=0.48]{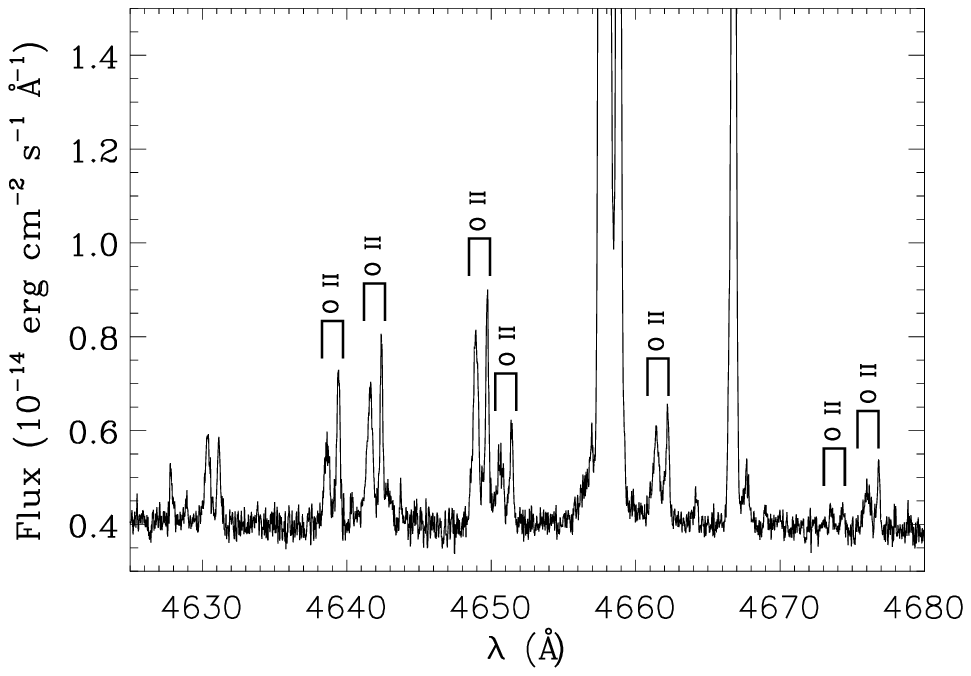} 
   \caption{Section of the echelle spectrum of \hh-S showing the shock (left) and 
            nebular (right) components of each emission line of multiplet 1 of \ion{O}{ii}.}
   \label{espec}
  \end{figure}  
 identification process and the measurement of line fluxes were done simultaneously. The 
 inspection of the line shapes at the bi-dimensional echelle spectrum was always used to identify 
 which component --nebular or shock-- was measured at each moment. The rather different spatial 
 and spatio-kinematic structure of the two kinematic components is illustrated in Figure~\ref{f2}. 
 We have identified 360 emission lines in the spectrum of \hh-S, 115 of them only show one 
 component --8 belong to the nebular component and 107 belong to the shock one-- and 8 are dubious 
 identifications.\\
 For a given line, the observed wavelength is determined by the centroid of the Gaussian fit to 
 the line profile. For lines measured in different orders and/or spectral ranges, the average of 
 the different wavelength determinations has been adopted. From the adopted wavelength, the 
 heliocentric velocity, \vhel, has been calculated using the heliocentric correction appropriate 
 for the coordinates of the object and the moment of observation. The typical error in the 
 heliocentric velocity measured is about 1-2 \kms.\\
 All line fluxes with respect to H$\beta$, $F(\lambda)/F({\rm H}\beta)$, were dereddened using the 
 typical relation,
 \begin{equation}
   \frac{I(\lambda)}{I({\rm H}\beta)} = \frac{F(\lambda)}{F({\rm H}\beta)}
                                         10^{c({\rm H}\beta)f(\lambda)}. 
 \end{equation} 
 The reddening coefficient, c(H$\beta$), respresents the amount of interstellar extinction which is 
 the logarithmic extinction at H$\beta$, while $f(\lambda)$ is the adopted extinction curve 
 normalized to $f({\rm H}\beta) = 0$. The reddening coefficient was determined from the comparison 
 of the observed flux ratio of Balmer and Paschen lines --those not contaminated by telluric or 
 other nebular emissions-- with respect to H$\beta$ and the theoretical ones computed by 
 \cite{storeyhummer95} for the physical conditions of \te\ $=$ 10000 K and \nel\ $=$ 1000 \cmc. As 
 in Paper I, we have used the reddening function, $f(\lambda)$, normalized to H$\beta$ derived by 
 \cite{blagraveetal07} for the Orion Nebula. The use of this extinction law instead of the 
 classical one  \citep{costeropeimbert70} produces slightly higher c(H$\beta$) values and also 
 slightly different dereddened line fluxes depending on the spectral range (see Paper I). The final 
 c(H$\beta$) values obtained for the two kinematic components were weighted averages of the values 
 obtained for the individual lines: c(H$\beta$){\neb} $=$ 0.41$\pm$0.02 and c(H$\beta$){\sho} $=$ 
 0.45$\pm$0.02. Although not all the c(H$\beta$) values are consistent with each other (see 
 \S\ref{difechb}), the average values obtained are quite similar and consistent within the 
 uncertainties.\\
 We can compare the reddening values with those obtained from integral field spectroscopy data 
 presented in Paper I in the same area of \hh-S (see Figure~\ref{f1}) and corresponding to the 
 section $\Delta\alpha$ $=$ $[-4,-6]$ and $\Delta\delta$ $=$ $[-8,-5]$ (see Figure 3 of Paper I). 
 The average c(H$\beta$) in this zone is 0.65$\pm$0.15, which is higher than those determined for 
 UVES data. However, if we re-calculate the value of c(H$\beta$) from the UVES data using the 
 same Balmer lines as in Paper I, we obtain a value 0.5$\pm$0.1 in both kinematic components, a 
 value consistent with the PMAS one within the errors. These differences can be 
 related to several systematical disagreements found between the c(H$\beta$) values obtained from 
 different individual Balmer or Paschen lines (see \S\ref{difechb}).\\
 In the most complete work on the reddening distribution across the Orion Nebula, 
 \cite{odellyusefzadeh00} obtain values of c(H$\beta$) between 0.2 and 0.4 in the zone around 
 \hh-S, somewhat lower than our reddening determinations. This can be due to the fact that 
 \citeauthor{odellyusefzadeh00} use the extinction law by \cite{costeropeimbert70} which, as we 
 discuss in Paper I, produces lower c(H$\beta$) values than the more recent extinction law 
 \citep{blagraveetal07}. We have also re-calculated c(H$\beta$) from our UVES spectra making use 
 of the \citeauthor{costeropeimbert70} law, and we obtain values about 0.3, being now in agreement 
 with the determinations of \cite{odellyusefzadeh00}.\\
 In Table~\ref{lines}, the final list of line identifications (columns 1--3), $f(\lambda)$ values 
 (column 4), heliocentric velocities (columns 5 and 8) and dereddened flux line ratios (columns 6 
 and 9) for the nebular and shock component are presented. The observational errors associated with 
 the line dereddened fluxes with respect to H$\beta$ --in percentage-- are also presented in 
 columns (7) and (10) of Table~\ref{lines}. These errors include the uncertainties in line flux 
 measurement, flux calibration and error propagation in the reddening coefficient.\\
 In column (11) of Table~\ref{lines}, we present the shock-to-nebular line flux ratio for those 
 lines in which both kinematic components have been measured. This ratio is defined as:
 \begin{equation}
   \frac{I_{\rm sh}}{I_{\rm neb}} = \frac{[I(\lambda)/I({\rm H}\beta)]_{\rm sh}}%
                                {[I(\lambda)/I({\rm H}\beta)]_{\rm neb} } %
	                      = \frac{I(\lambda)_{\rm sh}}{I(\lambda)_{\rm neb}} \times %
				    \frac{I({\rm H}\beta)_{\rm neb}}{I({\rm H}\beta)_{\rm sh}}     
 \end{equation}
 where the integrated dereddened H$\beta$ fluxes are $I$(H$\beta$){\neb} $=$ (3.80$\pm$0.20)%
 $\times$10$^{-12}$ erg~cm$^{-2}$~s$^{-1}$ and $I$(H$\beta$){\sho} $=$ (6.00$\pm$0.20)%
 $\times$10$^{-12}$ erg~cm$^{-2}$~s$^{-1}$. The $I$\neb/$I$\sho\ ratios depend on each particular 
 line. In general, they are close to 1 for \ion{H}{i} lines but become less than 1 for higher 
 ionized species --except Fe ions--, and are typically greater than 1 for neutral species. A more 
 extensive discussion on this particular issue will be presented in section \S\ref{ionis}.\\
 In Figure~\ref{espec}, we show a section of our flux-calibrated echelle spectra around the lines 
 of multiplet 1 of \ion{O}{ii}. It can be seen that both the nebular and the shock components are 
 well separated and show a remarkable high signal-to-noise ratio.      
 \input{tablalines}
\section{Results} \label{resul}     
 \subsection{Excitation mechanism of the ionized gas in \hh-S}\label{excitation}
  In their recent work, \cite{odellhenney08} argue that the presence of a variety of ionization 
  stages in the ionized gas of \hh\ indicates that the flow also contains neutral material. They 
  interpret that fact as due to the impact of the flow with preexisting neutral material --perhaps 
  of the foreground veil-- or that the flow compresses the ambient ionized gas inside the nebula to 
  such degree that it traps the ionization front. Our results provide some clues that can help to 
  ascertain this issue. The value of some emission line ratios are good indicators of the presence 
  of shock excitation in ionized gas, especially [\ion{S}{ii}]/H$\alpha$ and 
  [\ion{O}{i}]/H$\alpha$. In our spectra, we find log([\ion{S}{ii}] 6717+31/H$\alpha$) values 
  which are almost identical in both kinematic components ($-$1.49 and $-$1.44 for the nebular and 
  shock component, respectively). These values are completely consistent with those expected for 
  photoionized nebulae and far from the range of values between $-$0.5 and 0.5, which is the 
  typical of supernova remnants and HH objects \citep[see Figure 10 of][]{rieraetal89}. On the 
  other hand, the values of log([\ion{O}{i}] 6300/H$\alpha$) that we obtain for the nebular and 
  shock component are of $-$2.66 and $-$2.22, somewhat different in this case, but also far from 
  the values expected in the case of substantial contribution of shock excitation 
  \citep{hartiganetal87}. Finally, we have also used the diagnostic diagrams of \cite{ragaetal08} 
  where the [\ion{N}{ii}] 6548/H$\alpha$ and [\ion{S}{ii}] 6717+31/H$\alpha$ $vs.$ [\ion{O}{iii}] 
  5007/H$\alpha$ ratios of \hh\ are found in the zone dominated by photoionized shocks. Therefore, 
  the spectrum of \hh-S seems to be consistent with the picture that the bulk of the emission in 
  this area is produced by photoionization acting on compressed ambient gas that has trapped the 
  ionization front inside the ionized bubble of the nebula. In the rest of the paper we will 
  provide and discuss further indications that \hh-S contains an ionization front.  
 \subsection{Physical Conditions} \label{scond}
  We have computed physical conditions of the two kinematic components using several ratios of 
  CELs following the same methodology as in Paper I and in \cite{mesadelgadoetal08a}. The 
  electron temperatures, \te, and densities, \nel, are presented in Table~\ref{cond}.\\
  We have determined \nel\ from [\ion{O}{ii}], [\ion{S}{ii}], [\ion{Cl}{iii}] and [\ion{Ar}{iv}] 
  line ratios using the {\sc nebular} package \citep{shawdufour95}. In the case of the \nel\ 
  obtained from [\ion{Fe}{iii}] lines, we have used flux ratios of 31 and 12 lines for the shock 
  and nebular component, respectively, following the procedure described by 
  \cite{garciarojasetal06}. For the nebular component, we have adopted the average value of 
  \nel([\ion{O}{ii}]), \nel([\ion{S}{ii}]) and \nel([\ion{Cl}{iii}]) excluding 
  \nel([\ion{Fe}{iii}]) and \nel([\ion{Ar}{iv}]) due to their discrepant values and very large 
  uncertainties. For the shock component, we have adopted the average of \nel([\ion{O}{ii}]), 
  \nel([\ion{Cl}{iii}]) and \nel([\ion{Fe}{iii}]), while \nel([\ion{S}{ii}]) has not been included 
  because the [\ion{S}{ii}] line ratio is out of the range of validity of the indicator. As we can 
  see in Table~\ref{cond}, the density of the shock component ($\sim$17000 \cmc) is much higher 
  than the density of the nebular one ($\sim$3000 \cmc).\\
  However, the bulk of the emission of the nebular component might come from behind \hh, and 
  the electron density that we have found for that component might not be the true one of the 
  pre-shock gas. In fact, taking into account that the velocity of the gas flow is 89 \kms\ 
  \citep{odellhenney08} and the typical sound speed of an ionized gas is about 10-20 \kms, we have 
  adopted a Mach number, $M$, for \hh\ of about 5 and, thus, the shock compression ratio should be 
  $M^2\sim25$. Using the density of the shock component (see Table~\ref{cond}), we obtain a 
  pre-shock density $\sim 17430/25 \approx 700$ \cmc. This value is lower than the 2890 \cmc\ 
  determined for the nebular component. Therefore, it seems clear that the bulk of the nebular 
  component does not refer to the gas in the immediate vicinity of \hh\ as we have mentioned in 
  \S\ref{obsred}.\\ 
  Electron temperatures have been derived from the classical CEL ratios of [\ion{N}{ii}], 
  [\ion{O}{ii}], [\ion{S}{ii}], [\ion{O}{iii}], [\ion{S}{iii}] and [\ion{Ar}{iii}]. Under the 
  two-zone ionization scheme we have adopted \te([\ion{N}{ii}]) as representative for the low 
  ionization zone and \te([\ion{O}{iii}]) for the high ionization zone. We have also derived 
  \te(\ion{He}{i}) using the method of \cite{apeimbertetal02} and state-of-the-art atomic data 
  (see \S\ref{abrl}).\\
  We have compared these temperature determinations with those obtained from the integral field 
  unit (IFU) data presented in Paper I. We have determined the mean \te\ values of the spaxels of 
  the section of the FOV of the PMAS data that encompasses the area covered by our UVES spectrum, 
  finding $<$\te([\ion{O}{iii}])$>$ = 8760$\pm$260 K and $<$\te([\ion{N}{ii}])$>$ = 9730$\pm$590 K. 
  These values are in agreement within the errors with those obtained in this paper (see 
  Table~\ref{cond}). The average density from the IFU data, obtained from the [\ion{S}{ii}] 
  line ratio, is 7300$\pm$3000 \cmc, a value between the \nel\ adopted for each kinematic 
  component from the UVES data.\\
  As we can see in Table~\ref{cond}, the \te\ values are quite similar in both components with 
  differences of the order of a few 100 K. The temperatures derived from [\ion{N}{ii}] lines are 
  higher than those derived from [\ion{O}{iii}] lines, which is a typical result observed in 
  previous works on the Orion Nebula \citep[e.g.][]{mesadelgadoetal08a,rubinetal03}, as well as in 
  Paper I. This is a likely result of the ionization stratification in the nebula. It is 
  interesting to note that the difference between both temperatures is smaller in the case of the 
  shock component, in this case, all the emission comes from a --probably-- much narrower slab of 
  ionized gas.\\
  The relatively low uncertainties in the physical conditions are due to the high signal-to-noise 
  ratio of the emission lines used in the diagnostics. \cite{blagraveetal06} computed the physical 
  conditions for HH~529 and they obtained similar results --higher densities in the shock 
  component but similar temperatures in both components-- though with comparatively larger errors.  
  \begin{table}
   \centering
   \begin{minipage}{75mm}
   \caption{Physical Conditions.}
   \label{cond}
    \begin{tabular}{clcc}
     \hline
                                 &   & Nebular   & Shock \\
      \multicolumn{2}{c}{Indicator}  & Component & Component \\     
     \hline
      \input{tabla_cond}
     \hline
    \end{tabular} 
   \end{minipage}
  \end{table}
 \subsection{Ionic abundances from CELs} \label{abcel}  
  Ionic abundances of \ioni{N}{+}, \ioni{O}{+}, \ioni{O}{2+}, \ioni{Ne}{2+}, \ioni{S}{+}, 
  \ioni{S}{2+}, \ioni{Cl}{+}, \ioni{Cl}{2+},  \ioni{Ar}{2+} and \ioni{Ar}{3+} have been derived 
  from CELs under the two-zone scheme and \tf $=$ 0, using the {\sc nebular} package. All 
  abundances were calculated for the shock and nebular component, except for \ioni{Ar}{3+}, which 
  was not detected in the spectrum of the shock component. The atomic data for \ioni{Cl}{+} are not 
  implemented in the {\sc nebular} routines, so we have used an old version of the 5-level atom 
  program of \cite{shawdufour95} --{\sc fivel}-- that is described by \cite{derobertisetal87}. This 
  program uses the atomic data for this ion compiled by \cite{mendoza83}.\\
  We have also measured [\ion{Ca}{ii}], [\ion{Cr}{ii}], [\ion{Fe}{ii}], [\ion{Fe}{iii}], 
  [\ion{Fe}{iv}], [\ion{Ni}{ii}] and [\ion{Ni}{iii}] lines. The abundances of these ions are also 
  presented in Table~\ref{abioni}. They were computed assuming the appropriate temperature under 
  the two-zone scheme and the procedures indicated below. In addition and only in the shock 
  component, we have detected a substantial number of lines of other quite rare heavy-element ions 
  as [\ion{Cr}{iii}], [\ion{Co}{ii}], [\ion{Co}{iii}], [\ion{Ti}{iii}] and, possibly, 
  [\ion{Cr}{iv}], [\ion{Co}{iv}], [\ion{Mn}{ii}] and [\ion{V}{ii}]. Unfortunately, we cannot 
  derive abundances from these lines due to the lack of atomic data for these ions.\\
  Two [\ion{Ca}{ii}] lines at 7291 and 7324 \AA\ were detected in the shock component. In order to 
  derive the \ioni{Ca}{+} abundance, we solved a 5-level model atom using the single atomic dataset 
  available for this ion \citep{melendezetal07}. Note that this is the first determination of the 
  \ioni{Ca}{+} abundance in the Orion Nebula and this poses a lower limit to the gas-phase Ca/H 
  ratio in this object.\\
  Two and four [\ion{Cr}{ii}] lines were measured in the nebular and shock component, respectively,
  although those at 8309 and 8368 \AA\ are very faint. [\ion{Cr}{ii}] lines can be affected by 
  continuum or starlight fluorescence as is also the case for the [\ion{Fe}{ii}] and [\ion{Ni}{ii}] 
  lines. We have computed the \ioni{Cr}{+} abundances using a 180-level model atom that treat 
  continuum fluorescence excitation as in \cite{bautistaetal96} and includes the atomic data of 
  \cite{bautistaetal08}. In order to consider the continuum fluorescence excitation we have assumed 
  that the incident radiation field derives entirely from the dominant ionization star $\theta^1$ 
  Ori C. As \cite{bautistaetal96}, we have calculated a dilution factor assuming a $T_{eff}$ = 
  39000 K, $R_{star}$ = 9.0$R_\odot$ (see \S\ref{width}) and a distance to the Orion Nebula of 
  414 pc \citep{mentenetal07}. In Table~\ref{abioni}, we include the \ioni{Cr}{+}/\ioni{H}{+} ratio 
  for the nebular and shock component. This is the first estimation of the \ioni{Cr}{+} abundance 
  in the Orion Nebula.\\    
  Several [\ion{Fe}{ii}] lines have been detected in our spectra. As in the case of [\ion{Cr}{ii}] 
  lines, most of them are affected by continuum fluorescence \citep[see][]{rodriguez99,%
  verneretal00}. Following the same procedure as for \ioni{Cr}{+}, we considered a 159 model atom 
  in order to compute the \ioni{Fe}{+} abundances using the atomic data presented in 
  \cite{bautistapradhan98}.\\
  Many [\ion{Fe}{iii}] lines have been detected in the two kinematic components and their flux is 
  not affected by fluorescence. For the calculations of the \ioni{Fe}{2+}/\ioni{H}{+} ratio, we 
  have implemented a 34-level model atom that uses collision strengths taken from \cite{zhang96} 
  and the transition probabilities of \cite{quinet96} as well as the new transitions found by 
  \cite{johanssonetal00}. The average value of the \ioni{Fe}{2+} abundance has been obtained from 
  31 and 12 individual emission lines for the shock and nebular component, respectively.\\
  One [\ion{Fe}{iv}] line has been detected in the shock component at 6740 \AA. The \ioni{Fe}{3+}/%
  \begin{table}
   \centering
   \begin{minipage}{80mm}
     \caption{Ionic abundances and abundance discrepancy factors.}
     \label{abioni}
    \begin{tabular}{lcccc}
     \hline
      & \multicolumn{2}{c}{Nebular Component} & \multicolumn{2}{c}{Shock Component} \\
     \hline
       \input{tabla_abadf}
     \hline
    \end{tabular}
    \begin{description}
      \item[$^a$] In units of 12+log(\ioni{X}{+n}/\ioni{H}{+}) .  
      \item[$^b$] Average value from positions 8b and 11 of \cite{walteretal92}.
    \end{description}
   \end{minipage}
  \end{table}
  \ioni{H}{+} ratio has been derived using a 33-level model atom where all collision strengths are 
  those calculated by \cite{zhangpradhan97} and the transition probabilities recommended by 
  \cite{froesefischerrubin04}.\\
  Several [\ion{Ni}{ii}] lines have been measured in both kinematic components but they are 
  strongly affected by continuum fluorescence \citep[see][]{lucy95}. As for the \ioni{Cr}{+} and 
  \ioni{Fe}{+} ions, we have used a 76-level model that includes continuum fluorescence excitation 
  and the new collisional data of \cite{bautista04} in order to compute the \ioni{Ni}{+} 
  abundances.\\
  We have measured several [\ion{Ni}{iii}] lines in the shock and nebular components. These lines 
  are not expected to be affected by fluorescence. The \ioni{Ni}{2+}/\ioni{H}{+} ratio has been 
  derived using a 126-level model atom and the atomic data of \cite{bautista01}.\\  
  The final adopted values of the ionic abundances are listed in columns (1) and (3) of 
  Table~\ref{abioni} for the nebular and shock component, respectively. Columns (2) and (4) 
  correspond to the ionic abundances of both components assuming the presence of temperature 
  fluctuations (see \S\ref{adp}). In this table, we have also included the 
  \ioni{C}{2+}/\ioni{H}{+} ratio obtained from UV CELs by \cite{walteretal92}. We have taken the 
  average of the values corresponding to their slit positions 8b and 11, the nearest positions to 
  \hh. The uncertainties shown in the table are the quadratic sum of the independent 
  contributions of the error in the density, temperature and line fluxes.\\
  The abundance determinations presented in Table~\ref{abioni} show the following behaviour: 
  ionic abundances determined from CELs of once ionized species are always higher in the shock 
  than in the nebular component; the twice ionized species of elements lighter than Ne (included), 
  show lower abundances in the shock than in the nebular component; and the twice ionized species 
  of elements heavier than Ne show similar abundances in both components except for iron, chromium 
  and nickel abundances that show substantially larger abundances in the shock component, something 
  that can be explained if a significant dust destruction occurs in this component (see 
  \S\ref{dust}).\\
  Finally, we have compared our abundance determinations from UVES data with those obtained in the 
  \hh-S region from the IFU data presented in Paper I. Integrating the spaxels of the section of 
  the FOV indicated in \S\ref{lir}, we obtain 12+log(\ioni{O}{2+}/\ioni{H}{+}) = 8.18$\pm$0.07 
  from CELs, which is in good agreement with the numbers presented in Table~\ref{abioni} for both 
  kinematic components. On the other hand, the average value of the \ioni{O}{+} abundance from 
  CELs is 8.06$\pm$0.l4. However, considering the large density dependence of this ionic abundance, 
  we have recalculated the \ioni{O}{+}/\ioni{H}{+} ratio adopting the physical conditions measured 
  in the shock component from UVES, finding a value of 8.26$\pm$0.09, which is in better agreement 
  with the UVES determinations for the shock component, the brightest one in the [\ion{O}{ii}] line 
  emission.      
 \subsection{Ionic abundances from RLs} \label{abrl}
  We have measured several \ion{He}{i} emission lines in the spectra of \hh, both in the nebular 
  and in the shock components. These lines arise mainly from recombination, but they can be 
  affected by collisional excitation and self-absorption effects. We have used the effective 
  recombination coefficients of \citet{storeyhummer95} for \ion{H}{i} and those computed by 
  \citet{porteretal05}, with the interpolation formulae provided by \citet{porteretal07} for 
  \ion{He}{i}. The collisional contribution was estimated from \citet{saweyberrington93} and 
  \citet{kingdonferland95}, and the optical depth in the triplet lines was derived from the 
  computations by \citet{benjaminetal02}. We have determined the He$^+$/H$^+$ ratio from a 
  maximum likelihood method \citep[MLM, ][]{peimbertetal00, apeimbertetal02}.\\
  To self-consistently determine \nel(\ion{He}{i}), \te(\ion{He}{i}), \ioni{He}{+}/\ioni{H}{+} 
  and the optical depth in the \ion{He}{i} 3889 line, $\tau_{3889}$, we have used the 
  adopted density obtained from the CEL ratios for each component as \nel(\ion{He}{i}) (see 
  Table~\ref{cond}) and a set of 16 \ion{He}{i} lines (at 3614, 3819, 3889, 3965, 4026, 4121, 4388, 
  4471, 4713, 4922, 5016, 5048, 5876, 6678, 7065 and 7281 \AA). We have discarded the \ion{He}{i} 
  5048 \AA\ line in the nebular component because it is affected by charge transfer in the CCD. So, 
  for the nebular component of \hh, we have a total of 16 observational constraints (15 lines plus 
  \nel), and for the shock component, we have 17 observational constraints (16 lines plus \nel). 
  Finally, we have obtained the best value for the 3 unknowns and $t^2$ by minimizing $\chi^2$. The 
  final $\chi^2$ parameters we have obtained are 7.53 for the nebular component and 12.34 for the 
  shock component, which indicate very good fits, taking into account the degrees of freedom. The 
  final adopted value of the \ioni{He}{+}/\ioni{H}{+} ratio for each component is included in 
  Table~\ref{abioni}.\\
  We have detected \ion{C}{ii} lines of multiplets 2, 3, 4, 6 and 17.02. The brightest of these 
  lines is \ion{C}{ii} 4267 \AA, which belongs to multiplet 6 and can be used to derive a proper 
  \ioni{C}{2+}/\ioni{H}{+} ratio. The rest of the \ion{C}{ii} lines are affected by fluorescence, 
  like multiplets 2, 3 and 4 \cite[see][]{grandi76} or are very weak, as in the case of the line 
  multiplet 17.02 that has an uncertainty of 40\% in the line flux.\\ 
  We have derived the \ioni{O}{+}/\ioni{H}{+} ratio from RLs for the shock and nebular components. 
  The \ion{O}{i} lines of multiplet 1 are very weak and they are partially blended with bright 
  telluric emission. In order to obtain the best possible abundance determination, we have used 
  different lines for each component: \ion{O}{i} 7775 \AA\ for the nebular component and 
  \ion{O}{i} 7772 \AA\ for the shock one, these are precisely the lines least affected by 
  line-blending.\\
  The high signal-to-noise of the spectra allowed us to detect and measure seven lines of the 
  multiplet 1 of \ion{O}{ii} as we can see in Figure~\ref{espec}. These lines are affected by 
  non-local thermal equilibrium (NLTE) effects \citep{ruizetal03}, therefore to obtain a correct 
  \ioni{O}{2+} abundance it is necessary to observe the eight lines of the multiplet. However, 
  these effects are rather small in the Orion Nebula --as well as in the observed components-- due 
  to its relatively large density. Then, assuming LTE, the \ioni{O}{2+} abundance from RLs 
  has been calculated considering the abundances obtained from the flux of each line of 
  multiplet 1 and the abundance from the estimated total flux of the multiplet 
  \citep[see][]{estebanetal98}.\\
  The abundance determinations in Table~\ref{abioni} show that the \ioni{He}{+}/\ioni{H}{+}, 
  \ioni{C}{2+}/\ioni{H}{+} and \ioni{O}{2+}/\ioni{H}{+} ratios derived from RLs are always very 
  similar in both shock and nebular component. In the case of \ioni{O}{+} abundances, the nominal 
  values determined for both components seem to be somewhat different (about 0.24 dex), but they 
  are marginally in agreement considering the large uncertainties of this ion abundance.\\  
  As in the previous subsection, we have compared our abundance determinations from RLs with those 
  obtained for \hh-S in Paper I. From the data of Paper I, we obtain 
  12$+$log(\ioni{O}{2+}/\ioni{H}{+}) $=$ 8.39$\pm$0.13 and 12$+$log(\ioni{C}{2+}/\ioni{H}{+}) $=$ 
  8.29$\pm$0.11, values which are in good agreement with those obtained in this paper.   
 \subsection{Abundance discrepancy factors} \label{adfs}
  We have calculated ionic abundances from two kinds of lines --RLs and CELs-- for three ions: 
  \ioni{C}{2+}, \ioni{O}{+} and \ioni{O}{2+}. We present their values for the two components in 
  Table~\ref{abioni}. We have computed the ADF for these ions using the following definition: 
  \begin{equation}
    {\rm ADF(X^{+i}) = log\Big(\frac{X^{+i}}{H^+}\Big)_{RL} - %
                       log\Big(\frac{X^{+i}}{H^+}\Big)_{CEL}.}
  \end{equation}  
  In the case of the ADF(\ioni{C}{2+}) it can only be estimated for the nebular component and from 
  the comparison of our determination from RLs and those from CELs for nearby zones taken from the 
  literature. The value of the ADF(\ioni{C}{2+}) amounts to 0.45 dex.\\
  On the one hand, the ADF(\ioni{O}{+}) can also be estimated in our spectrum and shows values very 
  close to zero. However, these ADF values are rather uncertain. On the other hand, as we can see 
  in Table~\ref{abioni}, the \ioni{O}{2+} abundance from RLs is the same for both components while 
  that from CELs is lower in the shock component, probably because the recombination rate increases 
  in the shock one. This fact produces an \adfo\ about 0.2 dex higher in the shock component than 
  in the nebular one. This striking result will be discussed in section \S\ref{adp}.\\ 
  The values of the ADFs of \ioni{C}{2+} and \ioni{O}{2+} for the nebular component are in 
  good agreement with those obtained by \cite{estebanetal04} for a zone closer to the Trapezium 
  cluster than \hh. In the case of the ADF(\ioni{O}{+}), both determinations disagree, 
  \cite{estebanetal04} report a much larger value (0.39 dex).   
 \subsection{Total abundances} \label{totab}
  \begin{table*}
   \centering
   \begin{minipage}{100mm}
     \caption{Adopted ICF values.}
     \label{icf}
    \begin{tabular}{lcccccc}
     \hline
      & & \multicolumn{2}{c}{Nebular Component} & \multicolumn{2}{c}{Shock Component} \\
     \hline
      Elements&Unseen Ion&    \tf=0      &     \tf$>$0   &    \tf=0      &     \tf$>$0   \\      
     \hline
       \input{tabla_icf}
     \hline
    \end{tabular}
    \begin{description}  
      \item[$^a$] From photoionization models by \cite{garnettetal99}.   
      \item[$^b$] Mean of Orion Nebula models.
      \item[$^c$] From correlations obtained by \cite{martinhernandezetal02}. 
      \item[$^d$] From photoionization models by \cite{rodriguezrubin05}.    
    \end{description}
   \end{minipage}
  \end{table*}  
  In order to derive the total gaseous abundances of the different elements present in our 
  spectrum, we have to correct for the unseen ionization stages by using a set of ionization 
  correction factors (ICFs). The adopted ICF values are presented in Table~\ref{icf} and the total 
  abundances in Table~\ref{abtot}. As in the case of the ionic abundances from CELs, these tables  
  include values under the assumption of \tf $=$ 0 --columns (1) and (3)-- and under the 
  presence of temperature fluctuations (see \S\ref{adp}) --columns (2) and (4)--.\\
  The total helium abundance has been corrected for the presence of neutral helium using the 
  expression proposed by \cite{peimbertetal92} based on the similarity of the ionization potentials 
  of {\ioni{He}{0}} (24.6 eV) and \ioni{S}{+} (23.3 eV). 
  \begin{equation}
   {\rm \frac{He}{H} = \big( 1 + \frac{S^+}{S-S^+} \big) \times \frac{He^+}{H^+} = %
                                               ICF(He^0) \times \frac{He^+}{H^+} },
  \end{equation}
  For C we have adopted the ICF(\ioni{C}{+}) derived from photoionization models of 
  \cite{garnettetal99} for the shock and nebular component. In order to derive the total 
  abundance of nitrogen we have used the usual ICF:
  \begin{equation}
    {\rm  \frac{N}{H} = \frac{O^+ + O^{2+}}{O^+} \times \frac{N^+}{H^+} = %
                                     ICF(N^{2+}) \times \frac{N^+}{H^+}}.
  \end{equation}
  This expression gives very different values of the ICF(\ioni{N}{2+}) for both components due to 
  their rather different ionization degree.\\
  The total abundance of oxygen is calculated as the sum of \ioni{O}{+} and \ioni{O}{2+} 
  abundances. The absence of \ion{He}{ii} lines in the spectra and the similarity between the 
  ionization potentials of \ioni{He}{+} and \ioni{O}{2+}, implies the absence of \ioni{O}{3+}. 
  In Table~\ref{abtot} we present the O abundances from RLs and CELs.\\ 
  The only measurable CELs of Ne in the optical range are those of \ioni{Ne}{2+} but the fraction 
  of \ioni{Ne}{+} can be important in the nebula. We have adopted the usual expression 
  \citep{peimbertcostero69} to obtain the total Ne abundance:
  \begin{equation}
   {\rm  \frac{Ne}{H} = \frac{O^+ + O^{2+}}{O^{2+}} \times \frac{Ne^{2+}}{H^+} = %
                                        ICF(Ne^{+}) \times \frac{Ne^{2+}}{H^+}}.
  \end{equation} 
  We have measured CELs of two ionization stages of S: \ioni{S}{+} and \ioni{S}{2+}. Then, we have 
  used an ICF to take into account the presence of \ioni{S}{3+} \citep{stasinska78} which is based 
  on photoionization models of  \hii\ regions.
  \begin{equation*}
    {\rm  \frac{S}{H} = \Big( 1 - \big[\frac{O^+}{O^+ + O^{2+}} \big]^3 \Big)^{-1/3} %
                                                   \times \frac{S^+ + S^{2+}}{H^+} =}
  \end{equation*}
   \vspace{-0.4cm}
  \begin{equation}
    {\rm  = ICF(S^{3+}) \times \frac{S^+ + S^{2+}}{H^+}}.
  \end{equation}                  
  Following \cite{estebanetal98}, we expect that the amount of \ioni{Cl}{3+} is negligible in the 
  Orion Nebula. Therefore, the total abundance of chlorine is simply the sum of \ioni{Cl}{+} and 
  \ioni{Cl}{2+} abundances.\\
  For argon, we have determinations of \ioni{Ar}{2+} and \ioni{Ar}{3+} but some contribution of 
  \ioni{Ar}{+} is expected. In Table~\ref{icf} we present the values obtained from two ICF schemes: 
  one obtained from correlations between \ioni{N}{2+}/\ioni{N}{+} $vs.$ \ioni{Ar}{2+}/\ioni{Ar}{+} 
  from ISO observations of compact \hii\ regions by \cite{martinhernandezetal02} and another one 
  --following \cite{osterbrocketal92}-- derived as the mean of Orion Nebula models by 
  \cite{rubinetal91} and \cite{baldwinetal91}.\\
  We have measured lines of three ionization stages of iron in the shock component --\ioni{Fe}{+}, 
  \ioni{Fe}{2+} and \ioni{Fe}{3+}-- and two stages of ionization in the nebular component 
  --\ioni{Fe}{+} and \ioni{Fe}{2+}--. For the shock component, we can derive the total Fe abundance 
  from the sum of the three ionization stages. For the nebular component --and also for the shock 
  one in order to compare-- we have used an ICF scheme based on photoionization models of 
  \cite{rodriguezrubin05} to obtain the total Fe/H ratio using only the \ioni{Fe}{2+} abundances, 
  which is given by:
  \begin{equation}
    {\rm  \frac{Fe}{H} = 0.9 \times \Big( \frac{O^+}{O^{2+}} \Big)^{0.08} \times %
                         \frac{Fe^{2+}}{O^{+}} \times \frac{O}{H}}.
  \end{equation}
  Finally, there is no ICF available in the literature to correct for the presence of 
  \ioni{Ni}{3+} in order to calculate the total Ni abundance. Nevertheless, we have applied a first 
  order ICF scheme based on the similarity between the ionization potentials of \ioni{Ni}{3+} 
  (35.17 eV) and \ioni{O}{2+} (35.12 eV): 
  \begin{equation}
   {\rm \frac{Ni^{3+}}{Ni} = \frac{O^{2+}}{O}}.
  \end{equation}
  Therefore,
  \begin{equation}
    {\rm  \frac{Ni}{H} = \frac{O}{O^{+}} \times
                         \Big(\frac{Ni^{+}}{H^{+}} +\frac{Ni^{2+}}{H^{+}}\Big)}.
  \end{equation}                        
  In general, the total abundances shown in columns (1) and (3) of Table~\ref{abtot} are quite 
  similar for the shock and nebular component within the errors, except for the nickel and iron 
  abundances, which are much larger in the shock component --see section \S\ref{dust} for a 
  possible explanation. The set of abundances for the nebular component are in very good agreement 
  with previous results of \cite{estebanetal04}. We have also compared our Ni abundance values with 
  the previous determination of \cite{osterbrocketal92} finding that our Ni/H ratio for the nebular 
  component is an order of magnitud lower. This difference is due to the large uncertainties in the 
  atomic data used by those authors \cite[see][]{bautista01}. 
  \begin{table}
   \centering
   \begin{minipage}{80mm}
     \caption{Total abundances$^a$.}
     \label{abtot}
    \begin{tabular}{lcccc}
     \hline
      & \multicolumn{2}{c}{Nebular Component} & \multicolumn{2}{c}{Shock Component} \\
     \hline
                  &    \tf=0      &     \tf$>$0   &    \tf=0      &     \tf$>$0   \\      
     \hline
       \input{tabla_abtot}
     \hline
    \end{tabular}
    \begin{description}
      \item[$^a$] In units of 12+log(\ioni{X}{+n}/\ioni{H}{+}) .  
      \item[$^b$] Average value from positions 8b and 11 of \cite{walteretal92}.
      \item[$^c$] Value derived from RLs.
      \item[$^d$] Adopting the ICF from the mean of Orion Nebula models.
      \item[$^e$] Adopting the ICF from ISO observations \citep{martinhernandezetal02}. 
      \item[$^f$] From \ioni{Fe}{+}$+$\ioni{Fe}{2+}$+$\ioni{Fe}{3+}.
      \item[$^g$] Assuming the ICF of equation (6).
    \end{description}
   \end{minipage}
  \end{table}
\section{Discussion} \label{discu}
 \subsection{Differences between the c(H$\beta$) coefficient determined with different lines} 
  \label{difechb}
\begin{figure}
    \centering
    \includegraphics[scale=0.45]{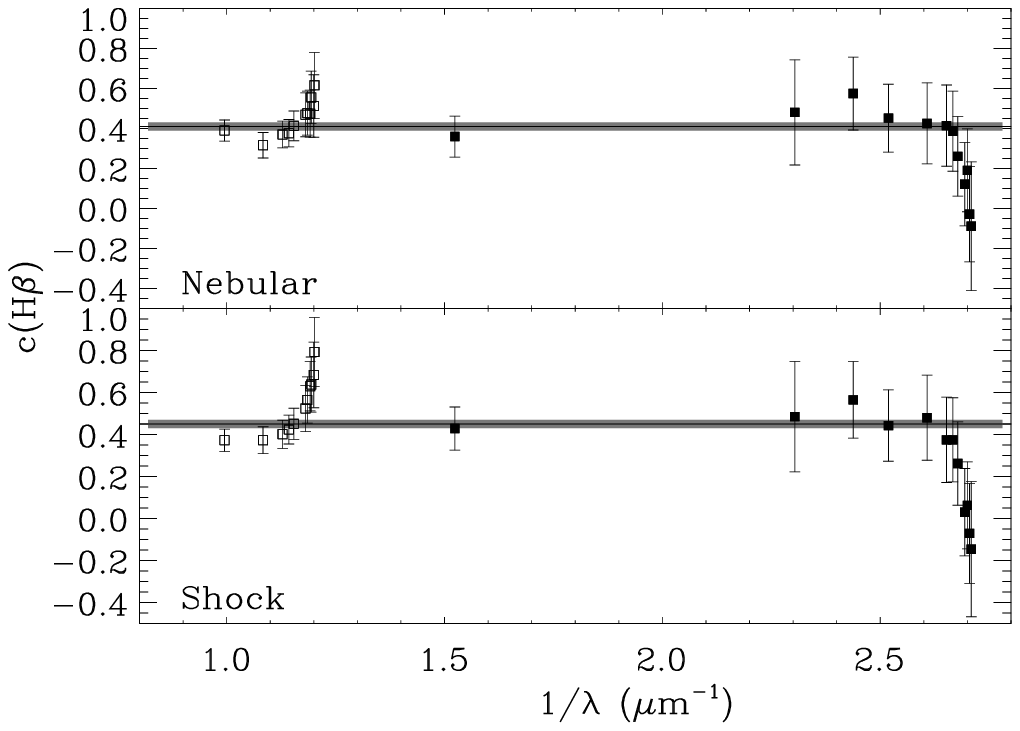} 
    \caption{c(H$\beta$) values obtained from different line ratios $vs.$ the inverse of the 
             wavelength in $\mu m$ for the nebular (up) and shock component (down). The filled 
	     and non-filled squares correspond to the c(H$\beta$) values derived from Balmer and 
	     Paschen lines, respectively. The horizontal line is the weighted average value adopted 
	     and the grey band its error}
    \label{chbvslambda}
  \end{figure}

\begin{figure}
    \centering
    \includegraphics[scale=0.45]{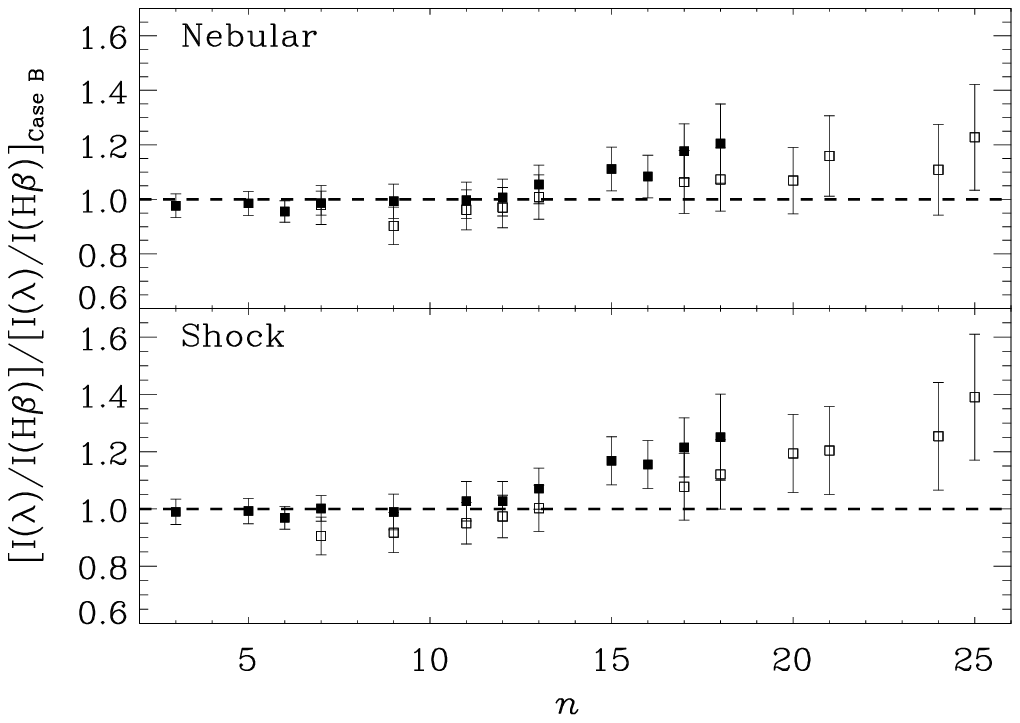} 
    \caption{Dereddened fluxes of Balmer (filled squares) and Paschen (non-filled squares) lines 
             to their theoretical flux under the Case B prediction ratio $vs.$ the principal 
	     quantum number $n$ for the nebular (up) and shock component (down).}
    \label{flujovsnivel}
  \end{figure}

  A puzzling feature of our UVES data is that the c$(\rm{H}\beta)$ values determined with different 
  lines ratios appear to be inconsistent with each other, even with observational errors taken into 
  account. Possible explanations are either a bias in the extinction curve or an extra mechanism, 
  in addition to extinction, altering the individual line intensities from case B predictions.\\
  Figure~\ref{chbvslambda} shows the c$(\rm{H}\beta)$ values measured for individual line ratios in 
  the Balmer and Paschen series. The bizarre pattern followed by the curve, and particularly the 
  steep slope it reaches in the proximity of the Balmer and Paschen limits, strongly suggest that 
  the solution cannot be a bias in the extinction law.\\
  We can test the second hypothesis by considering how the line intensity ratios relative to case B 
  ratios depend on the principal quantum number $n$ of the level where each line originates. Such 
  dependence, plotted in Figure~\ref{flujovsnivel} for the two kinematic components, shows a 
  definite trend with $n$ for both series. This strongly supports the second of our hypotheses, 
  namely that an extra mechanism is acting to deviate level populations away from case B 
  predictions. Indeed, detailed photoionization modeling indicates that this behaviour is the 
  result of two independent but concomitant mechanisms: $\ell$-changing collisions with 
  $|\Delta\ell|>1$ and pumping of Balmer and Paschen lines by absorption of stellar continuum 
  photons at the Lyman wavelengths. Both mechanisms are neglected in case B calculations 
  \citep{storeyhummer95} but included in our models, which take advantage of a new model hydrogen 
  atom with fully resolved levels (Cloudy, version C08.00; Porter, Ferland, van Hoof, \& Williams, 
  in preparation; see also Appendix A in Luridiana et~al. 2009); both alter the $n, \ell$ 
  populations, resulting in enhanced intensities of the high $n$ lines.\\ 
  As for the first mechanism, at low $n$ it has a negligible effect if compared to other 
  depopulation mechanisms, such as energy-changing collisions and horizontal collisions with 
  $|\Delta\ell| = 1$; at high $n$, it becomes increasingly important. Case B calculations neglect 
  this mechanism by construction, so a discrepancy is doomed to appear whenever high-$n$ lines are 
  compared to case B results.\\ 
  The effectiveness of the second mechanism strongly depends on the availability of the exciting 
  photons, $i.e.$ on the stellar flux at the Lyman wavelengths; the results of 
  \cite{Luridianaetal09} and further preliminary calculations (Luridiana et~al. in preparation) 
  suggest that its impact on line intensities might increase with $n$.\\ 
  A full account of both processes in \hii\ regions can be found in \cite{Luridianaetal09} and 
  Luridiana et~al. (in preparation).
 \subsection{Comparison of line ratios in the shock and nebular components and the ionization 
             structure} \label{ionis}
\begin{figure}
    \centering
    \includegraphics[scale=0.45]{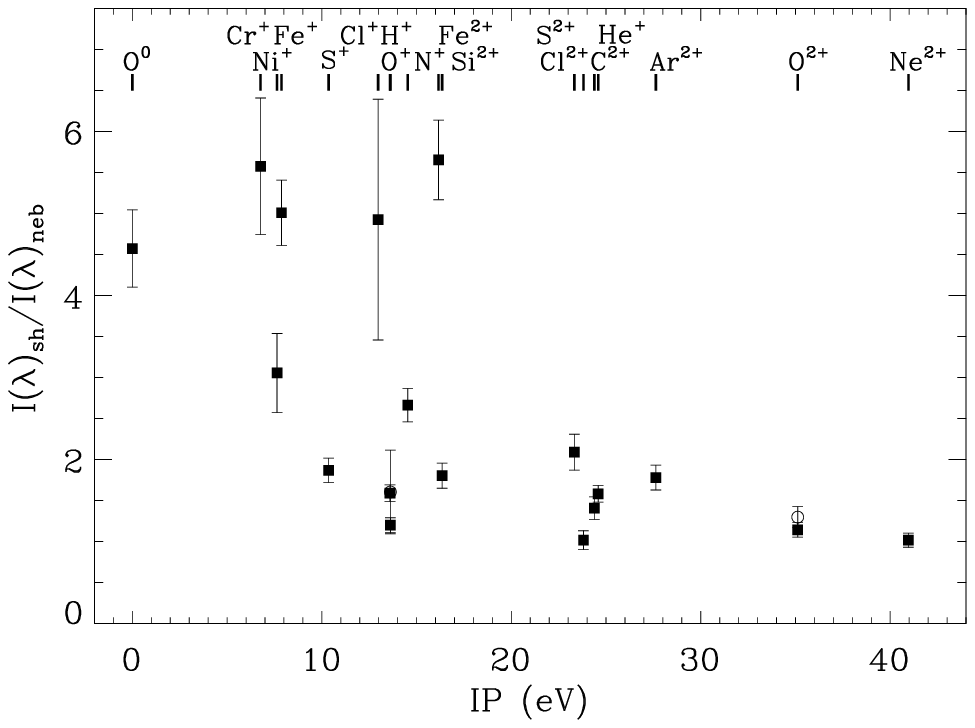} 
    \caption{Weighted average of shock-to-nebular line ratios $vs.$ the 
             ionization potencial (IP) needed to create the associated originating ion. Non-filled 
	     circles correspond to ratios measured from RLs (only for \ioni{O}{+} and 
	     \ioni{O}{++}). Smaller error bars are of about 3\%.}
    \label{coc}
  \end{figure}

  To maximize the shock-to-nebula ratio, the echelle spectra were extracted over the area where 
  the shock component is brighter and the velocity separation with respect to the nebular 
  background gas is maximum (see Figure~\ref{f2}).\\
  In Figure~\ref{coc}, we present the weighted average shock-to-nebular ratio for different ionic 
  species, $I$($\lambda$)\sho/$I$($\lambda$)\neb\ --which was defined in equation (1)-- with 
  respect to the ionization potential (IP) needed to create the associated originating ion. In 
  general, as we can see in this figure, the line ratios of the shock component relative to those 
  of the ambient gas are between 1 and 2 for most ionic species with IP above 10 eV, and close to 
  one for the most ionized species as \ioni{O}{2+} and \ioni{Ne}{2+}. Since the illumination of the 
  shock should be approximately the same as that of the nebula at this particular zone of the slit, 
  shock-to-nebular ratios of the order of one imply that the shock should be ionization-bounded. 
  This is exactly the opposite situation that \cite{blagraveetal06} find in HH~529, where the 
  shock-to-nebular ratios are clearly lower than 1 indicating that the shock associated to HH~529 
  is matter-bounded.\\
  As we have commented above and can be seen in Figure~\ref{coc}, the shock-to-nebular ratio varies 
  from 1 to 2 for species with an IP higher than 10 eV except in two cases: a) \ioni{Fe}{2+}, 
  whose lines show a substantial enhancement in the shock component, probably due to dust 
  destruction (see \S\ref{dust}), and b) \ioni{Cl}{+}, but this can be accidental because 
  this ionic abundance is derived from a single rather faint line with a very large uncertainty. 
  Ionic species with an IP lower than 10 eV show a shock-to-nebular ratio always higher than 2 in 
  Figure~\ref{coc}. In particular, \ioni{Fe}{+}, \ioni{Ni}{+} and \ioni{Cr}{+} show ratios larger 
  than 2 and these ions may be also affected by an increase of the gas-phase abundance due to dust 
  destruction.\\ 
  Neutral species like \ioni{O}{0} or \ioni{N}{0} are associated with the presence of an ionization 
  front. In the case of \ioni{O}{0}, the shock-to-nebular ratio has been calculated from the 
  [\ion{O}{i}] 6300 and 6363 \AA\ lines that are contaminated by telluric emissions but only in 
  the shock component. We extracted the telluric emissions from the zone free of shock emission 
  along the slit and subtracted this feature to the shock component. The shock-to-nebular ratio of 
  [\ion{O}{i}] lines is the largest one for those ions which are not heavily affected by possible 
  dust destruction. This is a further indication of the presence of an ionization front in \hh-S. 
  In the case of [\ion{N}{i}] lines, those belonging to the nebular component are also contaminated 
  by telluric emission but, in this case, it was impossible to deblend properly these lines.\\ 
  In Figure~\ref{f2} we can see the spatio-kinematic profiles of lines of different ionization 
  stages of oxygen: \ioni{O}{0}, \ioni{O}{+} and \ioni{O}{2+}, including RLs and CELs. In the shock 
  component, we can clearly distinguish a stratification in the location of the ions: the bulk of 
  the \ioni{O}{2+} emission is located at the south of the extracted area, \ioni{O}{0} at the north 
  and \ioni{O}{+} is located approximately at the centre of the extracted area. Another interesting 
  feature that can be seen in Figure~\ref{f2} is that the emissions from RLs and CELs of the same 
  ion seem to show the same spatio-kinematic profiles. These profiles provide further indication 
  that \hh-S is an ionization-bounded shock.\\
  The peak emission of the spatio-kinematic profiles of ions with IP lower or similar to the one 
  of \ioni{O}{+}:  \ioni{Cr}{+}, \ioni{Ni}{+}, \ioni{S}{+}, \ioni{Fe}{+} or \ioni{H}{+} is located 
  to the north of \hh-S, as in the case of \ioni{O}{+}. Ions as \ioni{He}{+}, \ioni{C}{2+} or 
  \ioni{Fe}{2+} show their peak emission about the centre of the aperture, while the ions with the 
  highest IP, as \ioni{Ne}{2+} or \ioni{Ar}{3+}, show spatio-kinematic profiles similar to 
  that of \ioni{O}{2+}.  
 \subsection{Width of the ionized slab and the physical separation between $\theta^1$Ori C and 
             \hh-S} \label{width} 
  An interesting result of this paper is the claim that \hh-S contains an ionization front as we 
  have shown in \S\ref{excitation} and \S\ref{ionis}. Due to this fact, we can estimate the width 
  of the ionized slab of \hh-S and its physical separation with respect to the main ionization 
  source of the Trapezium cluster, $\theta^1$Ori C.\\ 	     
  On the one hand, from the maximum emission of the shock component in the spatio-kinematic 
  profiles of \ioni{O}{2+} and \ioni{O}{0} shown in Figure~\ref{f2}, we can measure an angular 
  distance on the plane of the sky of about 3\farcs9$\pm$0\farcs5. Using the distance to the Orion 
  Nebula obtained by \cite{mentenetal07}, $d$ = 414$\pm$7 pc, and the inclination angle of \hh-S 
  with respect to the plane of the sky calculated by \cite{odellhenney08}, $\theta$ $=$ 48$^o$, we 
  estimate (11.7$\pm$1.5)$\times$10$^{-3}$ pc for the width of the ionized slab.\\
  On the other hand, to trap the ionization front in \hh-S, the incident Lyman continuum flux must 
  be balanced by the recombinations in the ionized slab, $i.e.$,
  \begin{equation}
    F_{\rm Ly} = \frac{Q(H^0)}{4\pi D^2} = n_{\rm sh}^2 \alpha_B(H^0,T) L,
  \end{equation}
  where $D$ is the physical separation between \hh-S and $\theta^1$Ori C, $Q(H^0)$ is the 
  ionizing photon rate, $n_{\rm sh}$ the density in the shock component, $\alpha_B(H^0,T)$ is the 
  case B recombination coefficient for H and $L$ the width of the slab. In order to estimate 
  $Q(H^0)$, we have used a spectral energy distribution of {\sc fastwind} code with the 
  stellar parameters for $\theta^1$Ori C obtained by \cite{simondiazetal06} 
  --$T_{eff} = 39000\pm1000$ K and log~$g = 4.1\pm0.1$ dex-- and the distance to the Orion Nebula 
  calculated by \cite{mentenetal07}. Then, the output parameters have been: the stellar radius 
  $R = (9.0\pm1.3) R_\odot$, the stellar luminosity log~$L = 38.80\pm0.14$ dex and the ionizing 
  photon rate $Q(H^0) = (6.30\pm2.00)\times 10^{48}$ s$^{-1}$. Taking a value for the recombination 
  coefficient, $\alpha_B = 2.59\times10^{-13}$ \cmc\ s$^{-1}$ at $10^4$ K, we have finally 
  calculated a physical separation of $D=0.14\pm0.05$ pc. This result suggests that \hh\ is quite 
  embedded within the body of the Orion Nebula and, therefore, discarts the origin of the ionized 
  front as result of the interaction of the gas flow with the veil (see \S\ref{excitation}), which 
  is between 1 and 3 pc in front of the Trapezium cluster \cite[see][]{abeletal04}.
 \subsection{Radial velocity analysis} \label{velocities}
\begin{figure}
    \centering
     \includegraphics[scale=0.45]{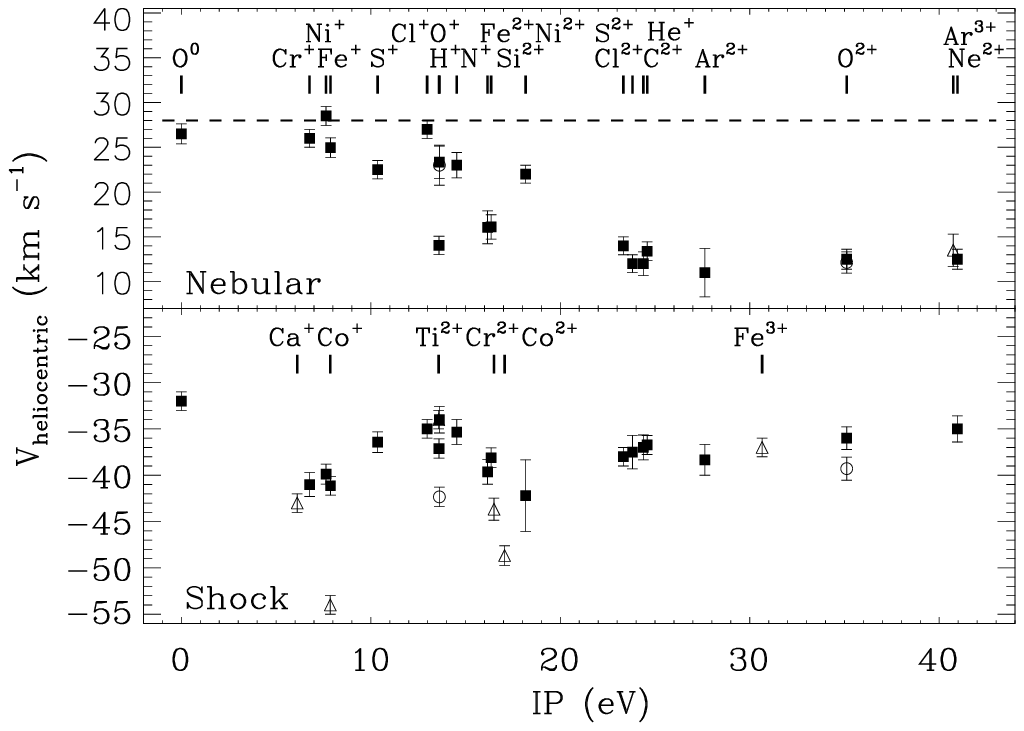} 
     \caption{Heliocentric velocities of the nebular (up) and shock (down) component emission lines 
              $vs.$ the ionization potencial (IP) needed to create the associated originating ion. 
	      The velocity of the photon dominated region (dashed horizontal line) is also shown. 
	      Non-filled triangles correspond to ions observed in only one kinematic component. 
	      Non-filled circles correspond to heliocentric velocities measured from RLs (only for 
	      \ioni{O}{+} and \ioni{O}{++}).}
    \label{vel}
   \end{figure}

  In Figure~\ref{vel}, we show the average heliocentric velocity of the lines that belong to a 
  given ionic species as a function of the IP needed to create the associated originating ion. We 
  have used two separate graphs to distinguish between the behaviour of the nebular and the shock 
  component. The nebular component shows the typical velocity gradient that has been observed in 
  other positions of the Orion Nebula and other \hii\ regions \citep[e.g.][]{bautistapradhan98,%
  estebanpeimbert99}, with a velocity difference of the order of $-$15 \kms\ between the neutral 
  and the most ionized species. This gradient is likely produced by the presence of flows of 
  ionized gas originating from the ionization front inside the nebula. In contrast, the ions in 
  the shock component show a rather similar radial velocity independently of their IP. This 
  indicates that the bulk of the ionized gas at the \hh-S is moving at approximately the same 
  velocity with respect to the rest of the nebula. The \hi\ lines (Balmer and Paschen series) of 
  the shock component are shifted by $-$50.7$\pm$1.0 \kms\ relative to the \hi\ lines of the 
  nebular component and $-$64.8$\pm$1.0 \kms\ relative to the velocity of the photon dominated 
  region \citep[PDR, +28 \kms;][]{goudis82}. The zone covered by our UVES slit (see 
  Figure~\ref{f1}) coincides with position 117$-$256 of \cite{doietal04}. These authors detect two 
  radial velocity components belonging to the shock gas in this zone, a fast component ($-$57 \kms) 
  and a slower brighter one ($-$31 \kms). Our value of the radial velocity for the shock 
  component of \hh\ ($-$36.8$\pm$1.0 \kms) is somewhat more negative than the slower velocity 
  component of \cite{doietal04}, probably our shock component corresponds to the unresolved blend 
  of the two velocity systems detected by those authors. 
 \subsection{The abundance discrepancy and temperature fluctuations} \label{adp}
  Assuming the validity of the temperature fluctuations paradigm and that this phenomenon produces 
  the abundance discrepancy, we can estimate the values of the \tf\ parameter from the ADFs 
  obtained for each component and ion (see Table~\ref{abioni}). In Table~\ref{t2}, we include the 
  \tf\ values that produce the agreement between the abundance determinations obtained from CELs 
  and RLs of \ioni{O}{2+} and \ioni{C}{2+}. These calculations have been made following the 
  formalism outlined by \cite{peimbertcostero69}. Adopting the \tf\ value obtained for 
  \ioni{O}{2+} zone, \tf(\ioni{O}{2+}), we have calculated the ionic abundances, the ICFs and the 
  total abundances under the presence of temperature fluctuations and they are presented in Tables 
  \ref{abioni}, \ref{icf} and \ref{abtot}, respectively.\\
  As we can see in Table~\ref{abtot}, the high \tf(\ioni{O}{2+}) value used to derive the 
  abundances for the shock component implies that the total abundances obtained from CELs 
  considering \tf$>$0 are higher in the shock component than in the nebular one for all the 
  elements. On the other hand, the abundances of C and O obtained from RLs are very similar in 
  both components. Moreover, the O/H ratio computed from CELs for \tf$>$0 is higher than that 
  obtained from RLs. However, for the nebular component, we find that the total abundance of 
  oxygen determined from CELs considering \tf$>$0 agrees with the oxygen abundance from RLs.\\ 
  These results can suggest that, perhaps, the \tf\ value we have found for the shock component 
  is too high and, therefore, that the \tf\ paradigm is not applicable in this case. However, we 
  have to consider that we have used a \tf\ value representative of the high ionization zone for 
  all the ions. The increase of the \ioni{O}{+}/\ioni{O}{2+} ratio in the shock component with 
  respect to the nebular one, due to an enhanced recombination rate, makes the \ioni{O}{+} zone 
  more extended in this component and a lower \tf(\ioni{O}{+}) would lead to a better agreement 
  between the O abundances of both components. We have also computed the \tf\ parameter for the 
  \ioni{He}{+} zone using a MLM (see~\S\ref{abrl}). The determination of the \tf(\ioni{He}{+}) 
  weighs the \ioni{O}{+} and the \ioni{O}{2+} zones depending on their extension. It is remarkable 
  that the \tf\ values obtained from different methods, such as the \ioni{He}{+} lines and the 
  \adfo, which assume the presence of temperature fluctuations in the observed volume, produce 
  almost identical values in both components, though \tf(\ioni{He}{+}) has large uncertainties. 
  These ones can reconcile the total abundances of both components for \tf$>$0 considering that 
  the MLM depends strongly on the \ion{He}{i} 3889 \AA\ line flux and that the shock components of 
  this line and H8 are severely blended.\\
  In Table~\ref{t2} we have also included the values of \tf\ obtained by \cite{blagraveetal06} 
  from the \adfo\ that they estimate for the nebular and shock components of HH~529 and those  
  obtained by \cite{estebanetal04} from the ADF of \ioni{O}{2+} and \ioni{C}{2+}. The data of 
  \cite{estebanetal04} correspond to a zone free of high velocity flows and can be considered as 
  representative of the nebular component but closer to the Trapezium. The values of the \tf\ 
  obtained for the nebular component of \hh-S are quite consistent with those obtained by 
  \cite{estebanetal04}. The \tf(\ioni{O}{2+}) of the nebular component of HH~529 
  obtained by \cite{blagraveetal06} is lower than the two other determinations, but still 
  consistent with our values within the uncertainties. As has been commented above, the \tf\ 
   \begin{table*}
    \centering
     \begin{minipage}{120mm}
      \caption{Estimates of the \tf\ parameter.}
      \label{t2}
      \begin{tabular}{lccccc}
       \hline
            &\multicolumn{2}{c}{\hh}&\multicolumn{2}{c}{HH~529$^a$}&\\
            &   Nebular &     Shock &   Nebular &     Shock &\\
        Ion & Component & Component & Component & Component &\cite{estebanetal04}\\     
       \hline
         \input{tabla_t2}
       \hline
      \end{tabular}
       \begin{description}
        \item[$^a$] \cite{blagraveetal06}.  
       \end{description}
     \end{minipage}
   \end{table*}
  values found in the shock component of \hh\ are much higher than those of the nebular component 
  and the value determined by \cite{blagraveetal06} for the shock component of HH~529.\\
  The effect of temperature fluctuations in the spectra of ionized nebulae and their existence has 
  been a controversial problem from the first work of \cite{peimbert67}. \cite{peimbert95}, 
  \cite{esteban02}, \cite{peimbertapeimbert06} have reviewed the possible mechanisms that could 
  produce such temperature fluctuations. Among the possible sources of temperature fluctuations, 
  we can list the deposition of mechanics energy by shocks and the gas compresion due to 
  turbulence. \cite{peimbertetal91} have studied the effect of shock waves in \hii\ regions using 
  shock models by \cite{hartiganetal87}. They found that high \tf\ values can be explained by the 
  presence of shocks with velocities larger than 100 \kms. In the case of the flow of 89 \kms\ that 
  produces \hh, this effect should produce very small \tf\ values. Therefore, another process or 
  other formalisms of the problem should be taken into account in this case. Further duscussion 
  of this problem is beyond the scope of this article.
\subsection{Dust destruction} \label{dust}
  As indicated in \S\ref{ionis}, the emission lines of refractory elements as Fe, Ni or Cr are 
  much brighter in the gas flow than in the ambient gas. Moreover, the shock component shows 
  relatively bright [\ion{Ca}{ii}] lines which are not detected in the spectrum of the ambient 
  gas. On the other hand, in \S\ref{excitation} we have stated that \hh-S does not contain a 
  substantial contribution of shock excitation and, therefore, we need other mechanisms to explain 
  those abnormally high line fluxes. It is well known that Fe, Ni, Cr and Ca are expected to be 
  largely depleted in neutral and molecular interstellar clouds as well as in \hii\ regions. 
  However, theoretical studies have shown that fast shocks --as those typical in HH objects and 
  supernova remnants-- should efficiently destroy grains by thermal and nonthermal sputtering in 
  the gas behind the shock front and by grain-grain collisions \citep{mckeeetal84,jonesetal94,%
  mouritaniguchi00}. Several works have shown that some nonphotoionized HH flows show a decrease 
  in the amount of Fe depletion as determined from the analysis of [\ion{Fe}{ii}] lines 
  \citep{beck-winchatzetal96,bohmmatt01,nisinietal05}. On the other hand, HH~399 is the first 
  precedent of a fully ionized HH object where an overabundance in Fe was detected and this one 
  was related with dust destruction \citep{rodriguez02}.\\
  In Table~\ref{depletions} we compare the values of the C, O, Fe, Ni and \ioni{Cr}{+} abundances 
  and the Fe/Ni ratios of the Sun \citep{grevesseetal07} and the nebular and shock components for 
  \tf$=$0 and \tf$>$0. In the case of Fe and Ni, we can see that the difference between the 
  abundances of the shock and nebular components are of the order of 0.85 dex for \tf$=$0 and 1.00 
  dex for \tf$>$0. This result indicates that the gas-phase abundance of Fe and Ni increases by a 
  factor between 7 and 10 after the passage of the shock wave. The fact that  the Fe/Ni ratio is 
  the same in both components --because the increase of the gas-phase abundance of both elements is 
  the same-- and consistent with the solar Fe/Ni ratio within the errors, suggests that the 
  abundance pattern we see in \hh-S is the likely product of dust destruction. In fact, 
  observations of Galactic interstellar clouds indicate that Fe, Ni --as well as Cr-- have the 
  same dust-phase fraction \citep{savagesembach96a,jones00}. It is also remarkable that, although 
  it could be modulated by ionization effects, the behaviour of the \ioni{Cr}{+} abundance is also 
  consistent with the dust destruction scenario. The increase of the \ioni{Cr}{+}/\ioni{H}{+} ratio 
  between the nebular and shock components is identical to that of Fe and Ni.\\
  Considering the solar Fe/H ratio as the reference \citep[12$+$log(Fe$/$H) $=$ 7.45$\pm$0.05,][]%
  {grevesseetal07} --which is almost identical to the Fe/H ratio determined for the B-type stars of 
  the Orion association \citep[12$+$log(Fe$/$H) $=$ 7.44$\pm$0.04,][]{przybillaetal08}-- we 
  estimate that the Fe dust-phase abundance decreases by about 30\% for \tf$=$0 and about 53\% for 
  \tf$>$0 after the passage of the shock wave in \hh-S. This result is in good agreement with the 
  predictions of the models of \cite{jonesetal94} and \cite{mouritaniguchi00}. In particular, for 
  the velocity determined for \hh\ \citep[89 \kms,][]{odellhenney08}, \cite{jonesetal94} 
  obtain a level of destruction of iron dust particles of the order of 40\%. On the other hand, the 
  Fe gas-phase abundance we measure in \hh-S follows closely the empirical correlation obtained by 
  \cite{bohmmatt01} from [\ion{Fe}{ii}] and [\ion{Ca}{ii}] emission line fluxes of several 
  nonphotoionized HH flows. Further evidence that the dust is not completely destroyed in \hh\ is 
  the detection of 11.7 $\mu m$ emission coincident with the bright ionized gas around \hh-S 
  \citep{smithetal05}. As we can see in Table~\ref{depletions}, the depletion factors of Fe and Ni 
  in the nebular component are similar to those found in warm neutral interstellar environments and 
  those at \hh-S of the order of the depletions observed in the Galactic halo 
  \citep[see][and references therein]{weltyetal99}.\\
  In the cases of C and O, the effect of dust destruction in their gas-phase abundance is more 
  difficult to estimate. Firstly, these elements are far less depleted in dust grains than Fe or 
  Ni in neutral interstellar clouds and in \hii\ regions. Secondly, there is still a controversy 
  about the correct solar abundance of these two elements \citep[see][]{holweger01,grevesseetal07}. 
  Finally, chemical evolution models predict some increase in the C/H and O/H ratios in the 4.6 
  Gyr since the formation of the Sun \citep[0.28 and 0.13 dex: ][]{carigietal05}. All these 
  problems make impossible to estimate confident values of the depletion factors for C and O.\\ 
  Considering the data gathered in Table~\ref{depletions}, the C abundance is virtually the same 
  in the nebular and shock components, although any possible small difference may be washed out 
  due to the intrinsic relatively large error of the abundance determination of this element. In 
  any case, only a slight increase of the C gas-phase abundance would be expected after the 
  passage of the shock wave for a shock velocity of about 89 \kms\ \citep{jonesetal94}.\\ 
  The uncertainties in the determination of the O abundances are much lower than in the case of C 
  and its determination does not depend on the selection of an appropriate ICF scheme. In 
  Table~\ref{depletions}, we see an increase of 0.06 dex in the O abundance in the shock component 
  with respect to the nebular one. These results suggest a possible moderate decrease of the 
  depletion for this element in the shocked material. Very recent detailed determinations of the O 
  abundance of B-type stars in the Orion association (Sim\'on-D\'\i az, in preparation) 
  indicate that the mean O/H ratio in this zone is 8.76$\pm$0.04. In principle, one would expect 
  that this is a better reference for estimating the dust depletion in the Orion Nebula than the 
  solar one because it corresponds to the $actual$ O abundance at the same location. If we take 
  the B-type stars determination as a reference for the total O/H ratio, the amount of O depletion 
  in the ambient gas would be $-$0.17$\pm$0.06 dex and $-$0.11$\pm$0.06 dex in the gas flow. 
  Therefore, a 30\% of the O tied up into dust grains would be destroyed after the passage of the 
  shock front.\\ 
  There are two other alternative methods to estimate the oxygen depletion factor. The first one 
  can be drawn following \cite{estebanetal98}. This one is based on the fact that Mg, Si and Fe 
  form molecules with O which are trapped 
  \begin{table*}
   \centering
   \begin{minipage}{120mm}
     \caption{Comparison of abundances and depletion factors}
     \label{depletions}
    \begin{tabular}{lcccccc}
     \hline
      & \multicolumn{2}{c}{ } & \multicolumn{2}{c}{Nebular Component} & %
        \multicolumn{2}{c}{Shock Component} \\
      &  \multicolumn{2}{c}{Sun$^a$} & \tf$=0$ & \tf$>0$ & \tf$=0$ & \tf$>0$ \\ 
	
     \hline
       \input{tabla_dust}
     \hline
    \end{tabular}
    \begin{description}
      \item[$^a$] \cite{grevesseetal07}.
    \end{description}
   \end{minipage}
  \end{table*}
  in dust grains. We have used Mg, Si and Fe depletions in order to obtain the fraction of O 
  trapped in dust grains. These depletions are estimated considering the Orion gas abundances 
  of Mg and Si given by \cite{estebanetal98}, our Fe abundance for \tf$>$0, the O abundances 
  derived from RLs, the stellar abundances of the Orion association of Si from Sim\'on-D\'\i az 
  (7.14$\pm$0.10) and Mg and Fe from \cite{przybillaetal08}. By assuming that O is trapped in 
  olivine (Mg, Fe)$_2$SiO$_4$, pyroxene (Mg, Fe)SiO$_3$ and several oxides like MgO, Fe$_2$O$_3$ 
  and Fe$_3$O$_4$ \citep[see][and references therein]{savagesembach96b}, we have estimated an O 
  depletion factor of $-$0.10$\pm$0.04 dex for the nebular component. If we consider the older 
  stellar abundances of Si and Fe from \cite{cunhalambert94} and the Mg/Fe ratio from 
  \cite{grevesseetal07}, this value becomes $-$0.08$\pm$0.05, which is almost identical to that 
  computed from the new stellar Mg, Si and Fe abundances.\\
  The last method to obtain O depletion factors assumes that oxygen 
  and iron are destroyed in the same fraction. The fraction of iron dust particules is calculated 
  using the Fe abundances for \tf$>$0 presented in Table~\ref{depletions} and Fe abundance of 
  B-type stars of the Orion association. From this assumption, and taking into account the O 
  abundance from RLs, we have derived a depletion factor of
  \begin{table*}
   \begin{minipage}{120mm}
    \centering \caption{Input parameters for photoionization models.}
    \label{models}
    \begin{tabular}{lcccc}
     \hline
       Parameter & Nebular Model & Shock Model A & Shock Model B & Shock Model C \\     
     \hline
       \input{tabla_model}
     \hline
    \end{tabular}
   \end{minipage}
  \end{table*}   
  $-$0.11$^{+0.11}_{0.14}$ dex in the nebular component. Finally, we have calculated the average of 
  the values obtained from the three methods, finding that the O depletion factor of the ambient 
  gas is $-$0.12$\pm$0.03. In all cases, the depletion becomes larger by about 0.08 dex if we adopt 
  abundances for \tf=0. Aditionally, the depletion factor in the shock component is smaller than in 
  the ambient gas, probably due to gas destruction.
  \begin{table*}
   \begin{minipage}{170mm}
    \centering \caption{Results of photoionization models.}
    \label{constraints}
    \begin{tabular}{lcccccc}
     \hline
      Constraint & Nebular Obs. & Nebular Mod. & Shock Obs. & Shock Model A & Shock Model B & %
                   Shock Model C\\
     \hline
       \input{tabla_compobmod}

     \hline
    \end{tabular}
   \end{minipage}
  \end{table*}  
 \subsection{Photoionization models for \hh-S} \label{smodels}  
  To test the role of dust destruction both in the temperature structure observed in the shock 
  component and in the iron abundance in the gas, we have run some simple photoionization models 
  for the nebular and the shock components. The models were constructed using Cloudy 
  \citep{ferlandetal98}, version 07.02, and assuming a plane-parallel open geometry with density 
  equal to the value adopted from the observations (see Table~\ref{cond}). We used a WM-basic 
  \citep{pauldrachetal01} model stellar atmosphere, with $T_{eff}$ $=$ 39000 K and log~$g$ $=$ 
  4.0, values very similar to those derived by \citet{simondiazetal06} for $\theta^1$~Ori~C. The 
  number of ionizing photons entering the ionized slab was specified using the ionization 
  parameter, $U$, the ratio of hydrogen-ionizing photons to the hydrogen density. We changed the 
  value of this parameter till the degrees of ionization (given by O$^{+}$/O$^{2+}$) derived for 
  the models were close to the observed ones. The final values used for $U$ imply similar numbers 
  of hydrogen ionizing photons in the nebular and shock models, with a difference of $\sim$40\%.\\
  We have used the Cloudy ``\hii\ region'' abundances, based on the abundances derived by 
  \citet{baldwinetal91}, \citet{rubinetal91} and \citet{osterbrocketal92} in the Orion Nebula, for 
  all elements except iron, which we have rescaled in order to reproduce the observed 
  [\ion{Fe}{iii}] 4658/H$\beta$ line ratios. The models also have ``Orion'' type dust: graphite 
  and silicate grains with the Orion size distribution \citep[deficient in small grains, see][]%
  {baldwinetal91} and an original dust to gas mass ratio of 0.0055. For each model, we used the 
  calculated line intensities to derive the physical conditions and the O$^{+}$ and O$^{2+}$ 
  abundances, following a similar procedure to the one used to derive the observational results.\\
  We computed three shock models. Model A has similar characteristics to the nebular model except 
  for the density, and can be used to assess the influence of this parameter on the electron 
  temperatures. Model B has the same input parameters as model A, but with the Fe abundance 
  multiplied by a factor of 7. Model C has the same input parameters as model B but with half the 
  amount of dust. The input model parameters are listed in Table~\ref{models}. Table~%
  \ref{constraints} shows a comparison for observations and models of the physical conditions, the 
  degree of ionization given by O$^{+}$/O$^{2+}$, and the  [\ion{Fe}{iii}] 4658/H$\beta$ line 
  ratio. We can see that the nebular model reproduces well the observational constraints. As for 
  the shock models, the increment in the density of model A is enough to explain the temperatures 
  found for the shock component, but does not reproduce the [\ion{Fe}{iii}] flux, whereas  the 
  higher Fe abundance of model B reproduces well this flux. Model C illustrates that a reduction 
  in the amount of dust does not change significantly the values of the chosen constraints. The 
  introduction of grains smaller than the ones considered in the Orion size distribution of Cloudy 
  will lead to higher temperatures through photoelectric heating, but we do not know what grain 
  size distribution would be suitable for the shock component.\\
  The iron abundances in those models that reproduce the observed [\ion{Fe}{iii}] line fluxes are 
  somewhat higher than the ones derived from the observations, but they reproduce the value of the 
  shock to nebular abundance ratio. This can be considered a confirmation of the increment in the 
  iron abundance in the shock component, which is most probably due to dust destruction. As it has 
  been discussed in \S\ref{dust}, dust 
  destruction could also increase the gaseous abundances of other elements like carbon or oxygen 
  and this increment will change the amount of cooling and hence the electron temperature. We ran 
  a model where the abundances of O and C were increased by 14\% and found that the derived 
  temperatures decrease by about 200 K.      
\section{Conclusions} \label{conclu}
 We have obtained deep echelle spectrophotometry of \hh-S, the brightest knot of the Herbig-Haro 
 object \hh. Our high-spectral resolution has permitted to separate two kinematic components: the 
 nebular component --associated with the ambient gas-- and the shock one --associated with the gas 
 flow. We have detected and measured 360 emission lines of which 352 lines were identified.\\
 We have found a clear disagreement between the individual c(H$\beta$) values obtained from 
 different Balmer and Paschen lines. We outline a possible solution for this problem based on 
 the effects of Ly-continuum pumping and l-changing collisions with protons.\\
 We have analyzed the ionization structure of \hh-S concluding that the dominant excitation 
 mechanism in \hh-S is photoionization. Moreover, the dependence of the $I$($\lambda$)\sho/%
 $I$($\lambda$)\neb\ ratios and the ionization potential, the comparison of the spatio-kinematic 
 profiles of the emission of different ions, as well as the physical separation estimated of 
 0.14$\pm$0.05 pc between \hh-S and $\theta^1$Ori C indicate that an ionization front is trapped 
 in \hh-S due to compression of the ambient gas by the shock.\\
 We have derived a high \nel, about 17000 \cmc, and similar \te\ for the low and high ionization 
 zones for the shock component, while for the ambient gas we obtain an \nel\ of about 3000 \cmc, 
 and a higher \te\ in the low ionization zone than in the high ionization one. We have estimated 
 that the pre-shock gas in the inmediate vicinity of \hh\ has a density of about 700 \cmc, 
 indicating that the bulk of the emission of the ambient gas comes from the background behind 
 \hh.\\
 We have derived chemical abundances for several ions and elements from the flux of collisionally 
 excited lines (CELs). In particular, we have determined the \ioni{Ca}{+} and \ioni{Cr}{+} 
 abundances for the first time in the Orion Nebula but only for the shock component. The abundance 
 of \ioni{C}{2+}, \ioni{O}{+} and \ioni{O}{2+} have been determined using recombination lines (RLs) 
 for both components. The abundance discrepancy factor for \ioni{O}{2+}, \adfo, is 0.35 dex in the 
 shock component and much lower in the ambient gas component. \\
 Assuming that the ADF and temperature fluctuations are related phenomena, we have found a 
 \tf(\ioni{O}{2+}) of 0.050 for the shock component and 0.016 for the nebular one. The high \tf\ 
 value of the shock component produces some apparent inconsistencies between the total abundances 
 in both components that cast some doubts on the suitability of the \tf\ paradigm, at least for the 
 shock component. However, the fact that the values of the \tf\ parameter determined from the 
 analysis of the \ion{He}{i} line ratios are in complete agreement with those obtained from the 
 \adfo\ supports that paradigm.\\ 
 Finally, the comparison of the abundance patterns of Fe and Ni in the nebular and shock 
 components and the results of photoionzation models of both components indicate that a partial 
 destruction of dust grains has been produced in \hh-S after the passage of the shock wave. We 
 estimate that the percentage of destruction of iron dust particles is of the order of 30\%-50\%. 
\section*{Acknowledgments}
 We are very grateful to the referee of the paper, J. Bally, for his comments, which have 
 improved the scientific content of the paper. We also thank S. Sim\'on-D{\'\i}az for providing us 
 the stellar parameters for $\theta^1$ Ori C. This work has been funded by the Spanish Ministerio 
 de Ciencia y Tecnolog\'\i a (MCyT) under project AYA2004-07466 and Ministerio de Educaci\'on y 
 Ciencia (MEC) under project AYA2007-63030. V.L. acknownleges support from MEC under project 
 AYA2007-64712. J.G.-R. is supported by a UNAM postdoctoral grant. M.R. acknowledges support from 
 Mexican CONACYT project 50359-F.


\label{lastpage}

\end{document}

%% file: tablalines.tex
\begin{table*}
 \centering
 \begin{minipage}{155mm}
  \caption{Identifications, reddening-corrected line ratios (I(H$\beta$)=100) for an area of 
           1\farcs5$\times$2\farcs5 and heliocentric velocities for the nebular and shock components.}
   \label{lines}	   
\begin{tabular}{clcccccccccc}
  \hline
         &  &  &  & \multicolumn{3}{c}{Nebular Component} & \multicolumn{3}{c}{Shock Component} &  \\
   $\lambda_0$ (\AA)$^a$ &  Ion$^a$ & Mult$^a$ & $f(\lambda)^b$ &  \vhel$^c$  & $I$(\ld)$^d$ & Error (\%)$^e$ & %
           \vhel$^c$& $I$(\ld)$^d$ & Error (\%)$^e$ & $I$\sho/$I$\neb$^f$ & Notes \\
 \hline
  \input{tabla1}
\end{tabular}
\end{minipage}
\end{table*}
\setcounter{table}{0}
\begin{table*}
 \centering
 \begin{minipage}{155mm}
  \caption{continued}
  \begin{tabular}{clcccccccccc}
  \hline
         &  &  &  & \multicolumn{3}{c}{Nebular Component} & \multicolumn{3}{c}{Shock Component} &   \\
   $\lambda_0$ (\AA)$^a$ &  Ion$^a$ & Mult$^a$ & $f(\lambda)^b$ &  \vhel$^c$  & $I$(\ld)$^d$ & Error (\%)$^e$ & %
           \vhel$^c$& $I$(\ld)$^d$ & Error (\%)$^e$ & $I$\sho/$I$\neb$^f$ & Notes \\
 \hline
  \input{tabla2}
\end{tabular}
\end{minipage}
\end{table*}
\setcounter{table}{0}
\begin{table*}
 \centering
 \begin{minipage}{155mm}
  \caption{continued}
  \begin{tabular}{clcccccccccc}
  \hline
         &  &  &  & \multicolumn{3}{c}{Nebular Component} & \multicolumn{3}{c}{Shock Component} &   \\
   $\lambda_0$ (\AA)$^a$ &  Ion$^a$ & Mult$^a$ & $f(\lambda)^b$ &  \vhel$^c$  & $I$(\ld)$^d$ & Error (\%)$^e$ & %
           \vhel$^c$& $I$(\ld)$^d$ & Error (\%)$^e$ & $I$\sho/$I$\neb$^f$ & Notes \\
 \hline
  \input{tabla3}
\end{tabular}
\end{minipage}
\end{table*}
\setcounter{table}{0}
\begin{table*}
 \centering
 \begin{minipage}{155mm}
  \caption{continued}
  \begin{tabular}{clcccccccccc}
  \hline
         &  &  &  & \multicolumn{3}{c}{Nebular Component} & \multicolumn{3}{c}{Shock Component} &   \\
   $\lambda_0$ (\AA)$^a$ &  Ion$^a$ & Mult$^a$ & $f(\lambda)^b$ &  \vhel$^c$  & $I$(\ld)$^d$ & Error (\%)$^e$ & %
           \vhel$^c$& $I$(\ld)$^d$ & Error (\%)$^e$ & $I$\sho/$I$\neb$^f$ & Notes \\
 \hline
  \input{tabla4}
\end{tabular}
\end{minipage}
\end{table*}
\setcounter{table}{0}
\begin{table*}
 \centering
 \begin{minipage}{155mm}
  \caption{continued}
  \begin{tabular}{clcccccccccc}
  \hline
         &  &  &  & \multicolumn{3}{c}{Nebular Component} & \multicolumn{3}{c}{Shock Component} &   \\
   $\lambda_0$ (\AA)$^a$ &  Ion$^a$ & Mult$^a$ & $f(\lambda)^b$ &  \vhel$^c$  & $I$(\ld)$^d$ & Error (\%)$^e$ & %
           \vhel$^c$& $I$(\ld)$^d$ & Error (\%)$^e$ & $I$\sho/$I$\neb$^f$ & Notes \\
 \hline
  \input{tabla5}
\end{tabular}
\end{minipage}
\end{table*}
\setcounter{table}{0}
\begin{table*}
 \centering
 \begin{minipage}{155mm}
  \caption{continued}
  \begin{tabular}{clcccccccccc}
  \hline
         &  &  &  & \multicolumn{3}{c}{Nebular Component} & \multicolumn{3}{c}{Shock Component} &   \\
   $\lambda_0$ (\AA)$^a$ &  Ion$^a$ & Mult$^a$ & $f(\lambda)^b$ &  \vhel$^c$  & $I$(\ld)$^d$ & Error (\%)$^e$ & %
           \vhel$^c$& $I$(\ld)$^d$ & Error (\%)$^e$ & $I$\sho/$I$\neb$^f$ & Notes \\
 \hline
  \input{tabla6}
\end{tabular}
\end{minipage}
\end{table*}
\setcounter{table}{0}
\begin{table*}
 \centering
 \begin{minipage}{155mm}
  \caption{continued}
  \begin{tabular}{clcccccccccc}
  \hline
         &  &  &  & \multicolumn{3}{c}{Nebular Component} & \multicolumn{3}{c}{Shock Component} &   \\
   $\lambda_0$ (\AA)$^a$ &  Ion$^a$ & Mult$^a$ & $f(\lambda)^b$ &  \vhel$^c$  & $I$(\ld)$^d$ & Error (\%)$^e$ & %
           \vhel$^c$& $I$(\ld)$^d$ & Error (\%)$^e$ & $I$\sho/$I$\neb$^f$ & Notes \\
 \hline
  \input{tabla7}
 \hline
\end{tabular}
  \begin{description}
   \item[$^a$] Identification of each line: laboratory wavelength, ion and multiplet.
   \item[$^b$] Value of the extinction curve adopted \citep{blagraveetal07}. 
   \item[$^c$] Heliocentric velocity in units of \kms, the typical error is 1-2 \kms.
   \item[$^d$] Dereddened fluxes with respect to $I$(H$\beta$) = 100.
   \item[$^e$] Error of the dereddened flux ratios. Colons indicate errors larger than 40 
               per cent.  
   \item[$^f$] Shock-to-nebular line flux ratio. See definition in equation (2).	        
   \item[$^g$] Line blended with another line and deblended via Gaussian fitting.
   \item[$^h$] Contaminated by ``ghost".
   \item[$^i$] Contaminated by telluric emissions and not deblended.
   \item[$^j$] Deblended from telluric emissions.     
  \end{description}
\end{minipage}
\end{table*}

%% file: tabla1.tex
 3187.74 &      \ion{He}{i} &            3 &   0.195 &      15 &   3.450 &  6 &     -34 &   3.682 &  6 &1.067 &   \\
 3239.74 &  [\ion{Fe}{iii}] &           6F &   0.194 &       - &       - &  - &     -35 &   0.806 & 15 &    - &   \\
 3286.19 &  [\ion{Fe}{iii}] &           6F &   0.192 &       - &       - &  - &     -35 &   0.206 & 15 &    - &   \\
 3319.21 &  [\ion{Fe}{iii}] &           6F &   0.191 &       - &       - &  - &     -38 &   0.114 & 40 &    - &   \\
 3322.54 &  [\ion{Fe}{iii}] &           5F &   0.191 &       - &       - &  - &     -48 &   0.744 &  9 &    - &   \\
 3334.90 &  [\ion{Fe}{iii}] &           6F &   0.190 &       - &       - &  - &     -36 &   0.320 &  9 &    - &   \\

%% file: tabla2.tex
 4009.22 &      \ion{He}{i} &           55 &   0.128 &      16 &   0.170 & 10 &     -35 &   0.222 & 10 &1.312 &$g$\\
 4023.98 &      \ion{He}{i} &           54 &   0.126 &      16 &   0.028 & 20 &     -39 &   0.026 & 20 &0.935 &   \\
 4026.08 &      \ion{N}{ii} &           40 &   0.126 &      13 &   2.195 &  6 &     -37 &   2.110 &  6 &0.961 &   \\
 4026.21 &      \ion{He}{i} &           18 &         &         &         &    &         &         &    &      &   \\
 4046.43 &  [\ion{Fe}{iii}] &           4F &   0.123 &       - &       - &  - &     -38 &   0.084 & 15 &    - &   \\
 4068.60 &    [\ion{S}{ii}] &           1F &   0.121 &      23 &   0.887 &  9 &     -35 &   5.318 &  6 &5.996 &   \\

%% file: tabla3.tex
 4471.47 &      \ion{He}{i} &           14 &   0.064 &      15 &   4.303 &  6 &     -35 &   4.405 &  6 &1.023 &   \\  
 4474.91 &   [\ion{Fe}{ii}] &           7F &   0.063 &      25 &   0.012 & :  &     -42 &   0.044 & 18 &3.655 &   \\
 4488.75 &   [\ion{Fe}{ii}] &           6F &   0.061 &       - &       - &  - &     -42 &   0.033 & 20 &    - &   \\
 4492.64 &   [\ion{Fe}{ii}] &           6F &   0.060 &      26 &   0.011 & :  &     -41 &   0.032 & 20 &3.025 &   \\
 4509.60 &   [\ion{Fe}{ii}] &           6F &   0.057 &       - &       - &  - &     -38 &   0.011 & :  &    - &$h$\\
 4514.90 &   [\ion{Fe}{ii}] &           6F &   0.056 &       - &       - &  - &     -41 &   0.024 & 40 &    - &$h$\\

%% file: tabla4.tex
 5055.98 &     \ion{Si}{ii} &            5 &  -0.030 &      17 &   0.224 &  9 &     -36 &   0.274 &  9 &1.223 &   \\
 5084.77 &  [\ion{Fe}{iii}] &           1F &  -0.034 &       - &       - &  - &     -40 &   0.332 &  9 &    - &   \\
 5111.63 &   [\ion{Fe}{ii}] &          19F &  -0.038 &       - &       - &  - &     -41 &   0.103 & 10 &    - &   \\
 5146.61 &       \ion{O}{i} &           28 &  -0.043 &      28 &   0.041 & 18 &       - &       - &  - &    - &   \\
 5146.65 &       \ion{O}{i} &           28 &         &         &         &    &         &         &    &      &   \\
 5158.00 &   [\ion{Fe}{ii}] &          18F &  -0.045 &       - &       - &  - &     -41 &   0.061 & 15 &    - &$g$\\

%% file: tabla5.tex
 6363.78 &     [\ion{O}{i}] &           1F &  -0.197 &      26 &   0.207 &  9 &     -32 &   0.554 &  9 &2.676 &$j$\\
 6371.36 &     \ion{Si}{ii} &            2 &  -0.198 &      15 &   0.107 & 15 &     -39 &   0.150 & 12 &1.406 &   \\
 6440.40 &   [\ion{Fe}{ii}] &          15F &  -0.206 &       - &       - &  - &     -41 &   0.037 & 18 &    - &   \\
 6533.80 &  [\ion{Ni}{iii}] &           2F &  -0.217 &       - &       - &  - &     -51 &   0.264 &  9 &    - &   \\
 6548.03 &    [\ion{N}{ii}] &           1F &  -0.218 &      24 &   15.60 &  5 &     -34 &   24.83 &  5 &1.591 &   \\
 6562.82 &       \ion{H}{i} &    H$\alpha$ &  -0.220 &      13 &   279.4 &  4 &     -38 &   283.1 &  4 &1.013 &   \\

%% file: tabla6.tex
 8267.94 &       \ion{H}{i} &          P34 &  -0.412 &      15 &   0.053 & 15 &     -37 &   0.067 & 11 &1.245 &   \\
 8271.93 &       \ion{H}{i} &          P33 &  -0.413 &      14 &   0.067 & 11 &     -39 &   0.068 & 11 &1.009 &   \\
 8276.31 &       \ion{H}{i} &          P32 &  -0.413 &      14 &   0.076 & 11 &     -38 &   0.086 & 11 &1.127 &   \\
 8281.12 &       \ion{H}{i} &          P31 &  -0.414 &      23 &   0.057 & 15 &     -36 &   0.104 & 11 &1.831 &  \\
 8286.43 &       \ion{H}{i} &          P30 &  -0.414 &      10 &   0.075 & 11 &     -46 &   0.042 & 15 &0.563 &$i$\\
 8292.31 &       \ion{H}{i} &          P29 &  -0.415 &      13 &   0.091 & 11 &     -37 &   0.111 & 11 &1.221 &  \\

%% file: tabla7.tex
10031.2  &      \ion{He}{i} &          7/7 &  -0.568 &      11 &   0.045 & 15 &     -38 &   0.054 & 11 &1.185 &   \\
10049.4  &       \ion{H}{i} &           P7 &  -0.569 &      13 &   5.425 &  7 &     -38 &   5.016 &  7 &0.924 &   \\
10320.5  &    [\ion{S}{ii}] &           3F &  -0.590 &      22 &   0.194 &  9 &     -37 &   1.016 &  7 &5.234 &   \\
10336.4  &    [\ion{S}{ii}] &           3F &  -0.591 &      22 &   0.216 &  9 &     -38 &   0.891 &  7 &4.117 &   \\
10370.5  &    [\ion{S}{ii}] &           3F &  -0.593 &       - &       - &  - &     -38 &   0.396 &  9 &    - &   \\

%% file: tabla_cond.tex
 n$_e$ (cm$^{-3}$) &  [\ion{O}{ii}] &    3490 $\pm$     810 &   18810 $\pm$    8280\\
                   &  [\ion{S}{ii}] &    2350 $\pm$     910 &  $>$14200            \\
                   &[\ion{Cl}{iii}] &    2470 $\pm$    1240 &   23780 $\pm$   13960\\
                   &[\ion{Fe}{iii}] &   11800 $\pm$    9000 &   17100 $\pm$    2500\\
		   & [\ion{Ar}{iv}] &    5800 :             &       -              \\
                   &        adopted &    2890 $\pm$     550 &   17430 $\pm$    2360\\
                   &                &                       &                      \\
         T$_e$ (K) &  [\ion{N}{ii}] &    9610 $\pm$     390 &    9240 $\pm$     300\\
                   &  [\ion{O}{ii}] &    8790 $\pm$     250 &    9250 $\pm$     280\\
                   &  [\ion{S}{ii}] &    8010 $\pm$     440 &    8250 $\pm$     540\\
                   & [\ion{O}{iii}] &    8180 $\pm$     200 &    8770 $\pm$     240\\
                   & [\ion{S}{iii}] &    8890 $\pm$     270 &    9280 $\pm$     300\\
                   &[\ion{Ar}{iii}] &    7920 $\pm$     450 &    8260 $\pm$     410\\
                   &    \ion{He}{i} &    8050 $\pm$     150 &    7950 $\pm$     200\\

%% file: tabla_abadf.tex
                   & \multicolumn{4}{c}{Ionic Abundances from CELs}\\     
   \hline
                  &    \tf=0      &     \tf$>$0   &    \tf=0      &     \tf$>$0   \\      
   \hline
      \ioni{C}{2+} & 7.87$^b$      &       -       &       -       &       -       \\
       \ioni{N}{+} & 7.02$\pm$0.04 & 7.07$\pm$0.05 & 7.35$\pm$0.03 & 7.52$\pm$0.04 \\
       \ioni{O}{+} & 8.00$\pm$0.06 & 8.05$\pm$0.09 & 8.29$\pm$0.06 & 8.48$\pm$0.08 \\
      \ioni{O}{2+} & 8.35$\pm$0.03 & 8.46$\pm$0.04 & 8.08$\pm$0.03 & 8.43$\pm$0.05 \\
     \ioni{Ne}{2+} & 7.46$\pm$0.11 & 7.58$\pm$0.12 & 7.13$\pm$0.10 & 7.51$\pm$0.11 \\
       \ioni{S}{+} & 5.50$\pm$0.07 & 5.54$\pm$0.08 & 6.03$\pm$0.04 & 6.22$\pm$0.05 \\
      \ioni{S}{2+} & 6.90$\pm$0.25 & 6.98$\pm$0.25 & 6.89$\pm$0.22 & 7.16$\pm$0.21 \\
      \ioni{Cl}{+} & 3.99$\pm$0.09 & 4.04$\pm$0.10 & 4.52$\pm$0.06 & 4.68$\pm$0.06 \\
     \ioni{Cl}{2+} & 5.13$\pm$0.04 & 5.23$\pm$0.05 & 5.05$\pm$0.05 & 5.38$\pm$0.06 \\
     \ioni{Ar}{2+} & 6.30$\pm$0.04 & 6.39$\pm$0.04 & 6.26$\pm$0.05 & 6.56$\pm$0.04 \\
     \ioni{Ar}{3+} & 3.73$\pm$0.11 & 3.85$\pm$0.12 &       -       &       -       \\
      \ioni{Ca}{+} &       -       &       -       & 3.86$\pm$0.07 & 4.03$\pm$0.07 \\
      \ioni{Cr}{+} & 2.88$\pm$0.11 & 2.92$\pm$0.11 & 3.75$\pm$0.07 & 3.91$\pm$0.07 \\
      \ioni{Fe}{+} & 5.18$\pm$0.26 & 5.23$\pm$0.27 & 5.82$\pm$0.03 & 6.01$\pm$0.06 \\
     \ioni{Fe}{2+} & 5.66$\pm$0.13 & 5.72$\pm$0.13 & 6.77$\pm$0.09 & 6.96$\pm$0.10 \\
     \ioni{Fe}{3+} &       -       &       -       & 5.87$\pm$0.16 & 6.16$\pm$0.20 \\
      \ioni{Ni}{+} & 3.83$\pm$0.10 & 3.88$\pm$0.11 & 4.78$\pm$0.09 & 4.96$\pm$0.09 \\
     \ioni{Ni}{2+} & 4.42$\pm$0.14 & 4.47$\pm$0.15 & 5.60$\pm$0.09 & 5.77$\pm$0.09 \\
   \hline
                   & \multicolumn{4}{c}{Ionic Abundances from RLs}\\     
   \hline
      \ioni{He}{+} & \multicolumn{2}{c}{10.94$\pm$0.01} & \multicolumn{2}{c}{10.93$\pm$0.01}\\
      \ioni{C}{2+} &  \multicolumn{2}{c}{8.32$\pm$0.07} &  \multicolumn{2}{c}{8.25$\pm$0.08}\\
       \ioni{O}{+} &  \multicolumn{2}{c}{8.01$\pm$0.12} &  \multicolumn{2}{c}{8.25$\pm$0.16}\\      
      \ioni{O}{2+} &  \multicolumn{2}{c}{8.46$\pm$0.03} &  \multicolumn{2}{c}{8.44$\pm$0.03}\\ 
   \hline  
                   & \multicolumn{4}{c}{ADFs}\\     
   \hline  
    \ioni{C}{2+} & \multicolumn{2}{c}{0.45}          &  \multicolumn{2}{c}{-}              \\
     \ioni{O}{+} & \multicolumn{2}{c}{0.01$\pm$0.17} &  \multicolumn{2}{c}{$-$0.04$\pm$0.14}\\
    \ioni{O}{2+} & \multicolumn{2}{c}{0.11$\pm$0.04} &  \multicolumn{2}{c}{  0.35$\pm$0.05}\\

%% file: tabla_icf.tex
      He &  \ioni{He}{0} & 1.04$\pm$0.02 & 1.03$\pm$0.02 & 1.12$\pm$ 0.06 & 1.08$\pm$0.04\\
  C$^a$  &   \ioni{C}{+} & \multicolumn{2}{c}{1.31$\pm$0.46} & \multicolumn{2}{c}{1.50$\pm$0.47}\\      
      N  &  \ioni{N}{2+} & 3.21$\pm$0.54 & 3.82$\pm$0.83 & 1.62$\pm$ 0.27 & 2.29$\pm$0.46\\
      Ne &  \ioni{Ne}{+} & 1.45$\pm$0.15 & 1.35$\pm$0.19 & 2.61$\pm$ 0.31 & 1.78$\pm$0.26\\
      S  &  \ioni{S}{3+} & 1.01$\pm$0.01 & 1.01$\pm$0.01 & 1.09$\pm$ 0.03 & 1.03$\pm$0.01\\
  Ar$^b$ &  \ioni{Ar}{+} & 1.33          &        -      &       -        &      -       \\
  Ar$^c$ &  \ioni{Ar}{+} & 1.20$\pm$0.36 & 1.16$\pm$0.36 & 2.00$\pm$ 0.51 & 1.71$\pm$0.46\\
  Fe$^d$ & \ioni{Fe}{3+} & 2.71$\pm$0.46 & 3.16$\pm$0.69 & 1.52$\pm$ 0.25 & 2.02$\pm$0.41\\
      Ni & \ioni{Ni}{3+} & 3.21$\pm$0.54 & 3.82$\pm$0.83 & 1.62$\pm$ 0.27 & 2.29$\pm$0.46\\

%% file: tabla_abtot.tex
      He & 10.95$\pm$0.01 & 10.95$\pm$0.02 & 10.98$\pm$0.03 & 10.98$\pm$0.02\\
   C$^b$ &  8.07          &        -       &        -       &        -      \\
   C$^c$ &  \multicolumn{2}{c}{8.43$\pm$0.17} &  \multicolumn{2}{c}{8.43$\pm$0.16}\\
       N &  7.53$\pm$0.08 &  7.62$\pm$0.11 &  7.56$\pm$0.08 &  7.81$\pm$0.10\\
       O &  8.51$\pm$0.03 &  8.60$\pm$0.04 &  8.50$\pm$0.04 &  8.76$\pm$0.05\\
   O$^c$ &  \multicolumn{2}{c}{8.59$\pm$0.05} & \multicolumn{2}{c}{8.65$\pm$0.05}\\
      Ne &  7.62$\pm$0.12 &  7.72$\pm$0.13 &  7.54$\pm$0.11 &  7.83$\pm$0.12\\
       S &  6.92$\pm$0.24 &  7.00$\pm$0.24 &  6.98$\pm$0.19 &  7.23$\pm$0.19\\
      Cl &  5.16$\pm$0.04 &  5.26$\pm$0.05 &  5.16$\pm$0.04 &  5.46$\pm$0.05\\
  Ar$^d$ &  6.42$\pm$0.04 &  6.52$\pm$0.04 &        -       &        -      \\
  Ar$^e$ &  6.38$\pm$0.19 &  6.46$\pm$0.14 &  6.56$\pm$0.21 &  6.79$\pm$0.12\\
  Fe$^f$ &       -        &       -        &  6.86$\pm$0.07 &  7.06$\pm$0.08\\
  Fe$^g$ &  6.10$\pm$0.15 &  6.19$\pm$0.16 &  6.95$\pm$0.12 &  7.19$\pm$0.13\\
      Ni &  5.03$\pm$0.14 &  5.12$\pm$0.15 &  5.87$\pm$0.11 &  6.11$\pm$0.12\\

%% file: tabla_t2.tex
 \ioni{O}{2+} & 0.016$\pm$0.006              & 0.050$\pm$0.007              %
              & 0.009$\pm$0.004              & 0.010$\pm$0.010              %
	      & 0.022$\pm$0.002\\
 \ioni{C}{2+} & 0.040                        &       -                      %
              &        -                     &        -                     %
	      & 0.039$\pm$0.011\\
 \ioni{He}{+} & 0.014$\pm$0.013              & 0.049$\pm$0.011              %
              &        -                     &         -              %
	      & -   \\

%% file: tabla_dust.tex
            C & \multicolumn{2}{c}{8.39$\pm$0.05} & \multicolumn{2}{c}{8.43$\pm$0.17} & \multicolumn{2}{c}{8.43$\pm$0.16} \\
            O & \multicolumn{2}{c}{8.66$\pm$0.05} & \multicolumn{2}{c}{8.59$\pm$0.05} & \multicolumn{2}{c}{8.65$\pm$0.05} \\
           Fe & \multicolumn{2}{c}{7.45$\pm$0.05} & 6.10$\pm$0.15     & 6.19$\pm$0.16 & 6.95$\pm$0.12     & 7.19$\pm$0.13 \\
           Ni & \multicolumn{2}{c}{6.23$\pm$0.04} & 5.03$\pm$0.14     & 5.12$\pm$0.15 & 5.87$\pm$0.11     & 6.11$\pm$0.12 \\
        Fe/Ni & \multicolumn{2}{c}{1.22$\pm$0.06} & 1.07$\pm$0.23     & 1.07$\pm$0.25 & 1.08$\pm$0.17     & 1.08$\pm$0.19 \\
 \ioni{Cr}{+} & \multicolumn{2}{c}{     -      }  & 2.88$\pm$0.11     & 2.92$\pm$0.11 & 3.75$\pm$0.07     & 3.91$\pm$0.07 \\
\hline
            & \multicolumn{2}{c}{Shock$-$Nebular} & \multicolumn{2}{c}{Nebular$-$Sun} & \multicolumn{2}{c}{Shock$-$Sun}\\
	      &    \tf$=0$        &   \tf$>0$     &    \tf$=0$        &    \tf$>0$    &    \tf$=0$        &    \tf$>0$   \\ 
\hline
            C & \multicolumn{2}{c}{0.00$\pm$0.27} & \multicolumn{2}{c}{0.04$\pm$0.19} & \multicolumn{2}{c}{0.04$\pm$0.18}\\
            O & \multicolumn{2}{c}{0.06$\pm$0.07} & \multicolumn{2}{c}{-0.07$\pm$0.07}& \multicolumn{2}{c}{-0.01$\pm$0.07}\\ 
           Fe & 0.85$\pm$0.13     & 1.00$\pm$0.14 &-1.35$\pm$0.05     &-1.26$\pm$0.05 &-0.50$\pm$0.06     &-0.26$\pm$0.09\\
           Ni & 0.84$\pm$0.12     & 0.99$\pm$0.13 &-1.20$\pm$0.04     &-1.11$\pm$0.04 &-0.36$\pm$0.06     &-0.12$\pm$0.10\\
        Fe/Ni &       -           &        -      &       -           &       -       &        -          &        -     \\
 \ioni{Cr}{+} & 0.87$\pm$0.07     & 0.99$\pm$0.07 &       -           &       -       &        -          &        -     \\

%% file: tabla_model.tex
log(\nel) (\cmc) & 3.46 	      & 4.24		   & 4.24		& 4.24               \\
log($U$)         & $-$1.95	      & $-$2.53 	   & $-$2.53		& $-$2.53            \\
Dust             & Orion	      & Orion		   & Orion		& 0.5$\times$Orion   \\
(Fe/H)$_{gas}$   & $2.1\times10^{-6}$ & $2.1\times10^{-6}$ & $1.5\times10^{-5}$ & $1.5\times10^{-5}$ \\

%% file: tabla_compobmod.tex
\nel\ (\cmc)                           & 2890$\pm$550     & 2960    & 17430$\pm$2360  & 17540	& 17670     & 17650\\
\te([\ion{N}{ii}]) (K)                  & 9610$\pm$390     & 9340    & 9240$\pm$300    & 9510	& 9220      & 9210\\
\te([\ion{O}{iii}]) (K)                 & 8180$\pm$200     & 8230    & 8770$\pm$240    & 8810	& 8670      & 8640\\
log(O$^+$/O$^{2+}$) 	                & $-$0.35$\pm$0.07 & $-$0.37 & 0.20$\pm$0.07   & 0.21   & 0.23	    & 0.22\\
$I$([\ion{Fe}{iii}] 4658)/$I$(H$\beta$) & 0.009$\pm$0.001  & 0.010   & 0.110$\pm$0.006 & 0.017	& 0.111     & 0.107\\